\documentclass[a4paper,12pt]{article}
\usepackage{hyperref}
\usepackage{color,xcolor}
\usepackage{amsmath,amsthm,amssymb}
\usepackage{graphicx}
\usepackage[authoryear]{natbib}

\newtheorem{thm}{Theorem}[section]

\newtheorem{lemma}[thm]{Lemma}
\newtheorem{proposition}[thm]{Proposition}
\theoremstyle{definition}

\theoremstyle{remark}
\newtheorem{remark}[thm]{Remark}
\newtheorem{exa}[thm]{Example}
\numberwithin{equation}{section}
\newcommand{\indicator}[1]{\mathbf{1}_{#1}}

\newcommand{\norm}[1]{\left\Vert#1\right\Vert}

\newcommand{\Real}{\bbR}
\newcommand{\Natural}{\bbN}

\newcommand{\inner}[2]{\big \langle #1 , #2 \big \rangle}
\newcommand{\prob}{\textsf{\upshape P}}
\newcommand{\qprob}{\textsf{\upshape Q}}
\newcommand{\expec}{\textsf{\upshape E}}

\newcommand{\dfn}{\, := \,}

\newcommand{\bbL}{\mathbb{L}}
\newcommand{\bbN}{\mathbb{N}}

\newcommand{\bbR}{\mathbb{R}}
\newcommand{\bbU}{\mathbb{U}}

\newcommand{\cF}{\mathcal{F}}
\newcommand{\cG}{\mathcal{G}}

\renewcommand{\d}{\mathrm{d}}

\newcommand{\MSE}{\mathbb{MSE}}
\newcommand{\DIS}{\mathbb{DIS}}

\newcommand{\id}{\mathsf{id}}
\newcommand{\trace}{\mathsf{tr}}

\newcommand{\market}{w}
\newcommand{\clo}{O}
\newcommand{\varnu}{\kappa}
\newcommand{\shrunk}{\rho}
\newcommand{\num}{\nu}
\newcommand{\numg}{\num}
\newcommand{\numf}{{\widehat{\num}}}
\newcommand{\funds}{f}
\newcommand{\multi}{\theta}

\mathchardef\mhyphen="2D
\newcommand{\EF}{\expec^{\cF}}
\newcommand{\PF}{\prob^{\cF}}

\begin{document}


\title{\bf Estimation of growth in fund models}

\author{ Constantinos Kardaras\footnote{E-mail: {\tt k.kardaras@lse.ac.uk}, Department of Statistics, London School of Economics, UK. } 
	\and
	Hyeng Keun Koo\footnote{E-mail: {\tt hkoo@ajou.ac.kr}, Department of Financial Engineering, School of Business, Ajou University, Korea. }
	\and
	Johannes Ruf\footnote{E-mail: {\tt j.ruf@lse.ac.uk}, Department of  Mathematics, London School of Economics, UK.}
}

\date{\today}

\maketitle \pagestyle{plain} \pagenumbering{arabic}

\abstract{
	Fund models are statistical descriptions of markets where all asset returns are spanned by the returns of a lower-dimensional collection of funds, modulo orthogonal noise. Equivalently, they may be characterised as models where the global growth-optimal portfolio only involves investment in the aforementioned funds.
	The loss of growth due to estimation error in fund models under local frequentist estimation is determined entirely by the number of funds. Furthermore, under a general filtering framework for Bayesian estimation, the loss of growth increases as the investment universe does. A shrinkage method that targets maximal growth with the least amount of deviation is proposed. Empirical evidence suggests that shrinkage gives a stable estimate that more closely follows growth potential than an unrestricted Bayesian estimate.
}

\begin{flushleft}
	{\footnotesize
		
	}
\end{flushleft}





\newpage
\section*{Introduction}

\subsection*{Discussion and contributions}

In this paper we study estimation of growth in continuous-time \emph{fund models}, where the returns of all available assets are spanned by the returns of a (typically, low-dimensional) subset of funds, up to an orthogonal noise residual.  We provide frequentist and Bayesian estimators of the {\em growth-optimal portfolio}, exploring theoretical underpinnings of the estimation, and investigate the expected  loss of growth due to estimation error. 

There has been a large and continuous accumulation of predictive signals (or characteristics) for asset returns, and currently the number of signals ranges from a few hundreds to several thousands---see \citet{GHZ2013}, \citet{HLZ2016}, \citet{MP2016}, \citet{YZ2017}, \citet{KNS2020}, and \citet{HXZ2020}. Modern computational and statistical techniques, including various machine learning methods, allow  to investigate the high-dimensional space of predictive signals and to make inferences on asset pricing. This stream of literature emphasizes dimension reduction\footnote{The dimension reduction problem in the face of a high-dimensional factor space is called the {\em multidimensional challenge} by \citet{Cochrane2011}.}, as well as the estimation and prediction of the maximum possible Sharpe ratio of portfolios (\citet{KPS2019}, \cite{KNS2020}, \citet{GKX2020,GKX2021}, \citet{GX}, \citet{GLX2021},  and \citet{Nagel2021}). The extant literature, however, does not consider the relationship between the dimension reduction and the estimation of the maximum Sharpe ratio. In this paper we investigate the relationship in the context of general continuous-time asset price dynamics. 
	
Our main contributions are fourfold. First, we provide a fully general characterisation of a factor pricing model in continuous time. 
	We show in Proposition \ref{prop:factor} that the growth optimal portfolio can be spanned by a set of funds if and only if the residuals of asset returns projected onto the returns of the funds are local martingales. This is an extension of the result in a static setting that asset returns have zero alphas when regressed onto a set of factors if and only if a stochastic discount factor can be constructed from the factors \citep[Section~6.3]{Cochrane2009}. Here we assume that the factors are indeed portfolios. This assumption is without loss of generality if the funds are interpreted as the  portfolios having the highest correlations with the  economic factors  in the intertemporal capital asset pricing model (ICAPM, \citet{Merton1973} and \citet{Breeden1979}).  The market model we consider, however, does not impose any assumptions other than general semimartingale continuous price processes. In particular, we do not assume any specific Markovian structure in which means and covariances of asset returns are described by a finite number of economic factors as in the ICAPM.\footnote{Hence our appellation of \emph{fund models} to avoid confusion with Markovian ``factor models''. }
The use of mimicking portfolios is prevalent in empirical studies of asset pricing, regardless of whether they are based on statistical factor models or an economic theory, e.g.,  arbitrage pricing theory,  ICAPM, or  $q$-theory (\citet{FF1993,FF2016}, \citet{DGTW1997}, \citet{MP2016}, \citet{KPS2019}, \citet{KNS2020}, \citet{HXZ2020}, and \citet{GKX2021}). The requirement that residuals are local martingales is similar to the zero alpha condition. 
	
	Second, we argue that the loss of growth arising from local frequentist estimation is proportional to the number of funds in the fund model, and does not depend on any other market characteristic. This is a universal property in the markets, previously not discussed in the literature. 
	In a general Bayesian setting, we also show that the loss of growth due to filtering, when restricted to investing in certain number of funds, is smaller if the fund model is correct than if it is misspecified.
	 Accordingly, dimension reduction and identification of a low-dimensional fund model is crucially important in reducing estimation errors. 
	 
	 Third, we provide a general framework for Bayesian estimation. 
	\citet{HZ1990}, \citet{Pastor2000}, \citet{PS2000},  \citet{PS2002},  \citet{Avramov2004}, \citet{AC2006}, \citet{BS2018}, and \citet{KNS2020} study Bayesian inference  on asset pricing models in a static setting.  We propose a general filtering problem which can provide a  framework for the Bayesian inference in a dynamic setting.  When filtering the theoretical model specification with the actual investors' information, important quantities in estimation and loss of growth are the first and second moments of the conditional law of the model growth-optimal portfolio. These two moments provide all the relevant information, therefore one does not need to actually calculate the whole posterior law.
	
	Fourth, we provide a shrinkage estimate of the growth optimal portfolio, which is practically important for  risk management. Recent discoveries document high estimates of the maximum Sharpe ratio, typically exceeding 2 (\citet{KPS2019}, \citet{GKX2020,GKX2021}).\footnote{The portfolios with high Sharpe ratios include large numbers of small stocks and large short positions; hence, they may not allow practical implementation, considering short-selling and transaction costs.} The high Sharpe ratio comes with high risk, since the volatility of the growth optimal portfolio is equal to the maximum Sharpe ratio. It would be a too aggressive strategy to manage a portfolio with volatility of its log return exceeding 200\%, even though it has a high expected return. We propose a shrinkage portfolio which is less risky than the unrestricted estimate of the growth optimal portfolio. Our empirical analysis with the US market shows that the shrinkage portfolio tracks the growth potential better and its risk is significantly smaller than the unrestricted one.  Our approach to shrinkage is different from the ones in the literature in which authors shrink the mean or the covariance towards the prior belief (\citet{Jorion1986}, \citet{BL1992}, \citet{LW2003},  \citet{Avramov2004}, and \citet{KNS2020}). We instead minimise the dispersion between the maximum  growth potential from the perspective of the investor and the actual  growth.

	   \bigskip

	Initially adopting a frequentist non-parametric approach, we derive the most efficient estimator of the growth-optimal portfolio. We show that the most efficient estimator uses only returns of the funds in the fund models;  any other cross-sectional data ``orthogonal" to the funds are \emph{irrelevant}. We also show that the estimation accuracy  increases as the total variance of the fund returns increases.

	We then investigate the expected loss of growth due to estimation error. Surprisingly, the expected loss depends \emph{only} on the number of funds, and not on any other characteristics of the market. In fact, the instantaneous expected loss of growth is simply linear in the number of funds spanning the asset returns.
	
	The frequentist  estimator is very noisy, due to the large error in the estimation of instantaneous returns. In order to have a practical and useful estimate, we move to a Bayesian formulation. There, even in fund models, other data (e.g., characteristics and macroeconomic data) may be important for estimation. For this purpose, we consider a filtering framework with two information flows, a model filtration and an observation filtration. The larger information flow  is used  as a modelling device and theoretically represents ``full'' information. The investor's coarser filtration corresponds to the available information flow, where we assume that at least the returns process is observable. A special case of our framework covers an investor's Bayesian prior on certain model parameters.  We show that important quantities for both estimation purposes and for calculation of the loss of growth  are the first and second moments of the conditional law of the theoretical model's growth-optimal portfolio; as these two moments give all the relevant information, the investor does not need to  calculate the whole posterior law.
	
	In the filtering problem, the loss of growth increases as the
	investment universe does, a result similar to that in the frequentist approach.  This, in particular, implies that the loss in growth due to filtering, when restricted to investing in certain number of funds, is smaller if the fund model is correct than if misspecified. The investor, however, may be under the impression that the large loss is due to estimation error than misspecification error.  
	
	We next consider an example in which the growth-optimal portfolio has a Gaussian prior, independent of the observation filtration, and derive its estimator in terms of cumulative return and covariance processes, which is more easily computable than the frequentist local estimates. By using an empirical Bayesian approach, we show that the loss of
	growth can be substantial, amounting to $5\%$ with a single-fund model
	and 10 years of prior observation.
	The large economic loss contrasts the small utility loss due to deviations from optimal behaviour in the presence of information or trading costs \citep{Cochrane1989}. 
	
	The optimal estimator based on filtering maximises the  growth rate among all the portfolios that can be formed using observable data.  This is  an aggressive strategy leading to maximal growth based on the available information. From the vantage point of an investor, however,  there may be a lot of “spread” in the conditional law  between the true growth rate and its expectation. It may be more appealing to take a slightly more conservative approach, and instead try to minimise the spread between the true growth rate and the maximal expected growth rate among all the portfolios constructed from available information,  thereby targetting maximal growth with the least amount of deviation from its true value, even if not fully achieving such maximal growth in expectation. We thus derive a \emph{best tracking of maximal growth} estimator, by using the first two moments of the conditional law of the model growth-optimal portfolio. The estimator  has a strong flavour of \emph{shrinkage}, in the sense that the resulting portfolio takes smaller positions in the risky asset; not only does it track growth better, but it also reduces the resulting wealth process variability.  We conduct an empirical analysis and show that this shrinkage method produces a stable portfolio, following the growth potential more closely compared to the unrestricted filtered estimate.

\subsection*{Related literature}

We briefly discuss related literature, in addition to the one mentioned previously. 

\citet{Barry1974}, \citet{KB1976}, and \citet{Brown1978} study the effect of the estimation error on optimal portfolios in a static setting under parameter uncertainty with a Bayesian approach. \citet{Barberis2000}  studies the effect in a multi-period discrete time setting. \citet{Williams1977}, \citet{Gennote1986}, \citet{Feldman1992}, \citet{Brennan1998}, and \citet{Xia2001} study the effect of parameter uncertainty and learning in a continuous time setting.  \citet{Brandt1999}, \citet{AB2001}, \citet{BGSS2005}, \citet{BSV2009} study the estimation of optimal portfolios in discrete time. In this literature, the authors estimate the optimal portfolio weights  from the first-order conditions implied by  utility maximisation   \citep[see ][for a review]{Brandt2010}. In contrast, in this paper we only consider the growth-optimal portfolio with a more comprehensive statistical approach. 
 There exists a vast literature on factor models. For a recent treatment of the
 topic we refer to \citet{FF2016}, \citet{HLZ2016}, \citet{FGX2020}, \citet{GKX2020}, \citet{CPZ2020}, \citet{GLX2021}, and \citet{GX}. To the best of our knowledge, the present paper is the first to investigate  estimation of growth in continuous-time fund models. 
 
 
Shrinkage estimation was developed by \citet{Stein1956}, \citet{JS1961}, and \citet{EM1973} to address the issue of non-efficiency of the traditional estimator of the mean in a multi-variate setting. \citet{JKR1979}, \citet{JK1981}, \citet{Jorion1986}, and \citet{KZ2007} study  shrinkage estimation of optimal portfolios in a static setting.   \citet{BL1992} adopt a Bayesian method to incorporate an investor's view, and propose an estimator that shrinks the view toward the market equilibrium.
 \citet{LW2003,LW2004}, \citet{JM2003}, and \citet{DGNU2009} study the shrinkage estimation of the covariance matrix of asset returns and its relationship with portfolio constraints. 
 Our approach is different; we derive the shrinkage estimator from the objective to minimise the spread between the true growth rate and the maximal expected growth rate, whereas the previous research derives it to minimise the mean squared error. 
 
 \citet{KS1996}, \citet{Avramov2004}, and \citet{KZ2007} investigate the utility loss of an estimated portfolio due to estimation error. The case when the coefficient of relative risk aversion equals  1 in these studies corresponds to the loss of growth in our paper.  In particular, \citet{KZ2007} show that the utility loss is proportional to the number of assets, similar to our result. They, however, consider the loss in a discrete-time  environment with independent and identically distributed shocks, without any consideration of asset pricing. We derive the result in a general semimartingale continuous-time market and study its effect in fund models.

We touch only tangentially the topic of estimation of stochastic processes in continuous time. An interested reader may consult \citet{AM2004}, \citet{A2009}, \citet{AHS2009}, and the references therein. 

\subsection*{Structure of the paper}

Section~\ref{S:Intro} describes the financial market and derives the growth-optimal portfolio. Section~\ref{S:FundModels} studies fund models, and the estimation of the growth-optimal portfolio under them. Section~\ref{S:Filtering} discusses estimation under the two different information structures and derives the economic loss due to coarser information. Section~\ref{S:Shrinkage} studies the shrinkage method. Section~\ref{S:empirics} contains an empirical study
and Section~\ref{S:conclusion} concludes.

\section{Market and growth optimality} \label{S:Intro}

We consider a financial market in continuous time, modelled under a stochastic environment. We use $(S_i; \, i \in I)$ to denote market prices 
of certain assets, already discounted by the (observable) short rate process. We shall be mainly concerned with long-term investment in this paper, which constitutes a rather macroscopic point of view; therefore, we assume that the prices processes are continuous. We consider dynamics of the form
\begin{equation} \label{eq:returns}
	\d R_i \dfn \frac{\d S_i}{S_i} = \d A_i + \d M_i, \qquad i \in I, 
\end{equation}
for the excess returns of companies, where $A \equiv (A_i; \, i \in I)$ are continuous and adapted finite variation processes, and $M \equiv (M_i; \, i \in I)$ are continuous local martingales. We write $\d C$ for the instantaneous increments of the covariation process of the returns, i.e.,
\[
\d C_{ij}  \dfn (\d R_i)(\d R_j)  = (\d \log S_i)  (\d \log S_j) = (\d  M_i)(\d  M_j) 
, \qquad (i, j) \in I \times I;
\]
the corresponding matrix-valued integrated process of quadratic covariations is $C \equiv (C_{ij}; \, (i, j) \in I \times I)$.

An investment in the portfolio process $\pi \equiv(\pi_i; \, i \in I)$, with the understanding that $\pi_i$ represents proportion of current wealth invested in the $i$th asset, leads to wealth process $X_\pi$ with dynamics\footnote{A ``$*$'' superscript denotes transposition throughout.}
\[
\frac{\d X_\pi }{X_\pi } = \sum_{j \in I} \pi_j \frac{\d S_j}{S_j} = \d R_\pi \dfn \pi^* \d A + \pi^* \d M.
\]
In log-wealth terms, an application of It\^o's formula gives
\[
\d \log X_\pi =  \frac{\d X_\pi}{X_\pi} -  \frac{1}{2} \left( \frac{\d X_\pi}{X_{\pi}}\right)^2 = \d \Gamma_\pi +   \pi^* \d M,
\]
where
\begin{equation}\label{eq:portfolio_growth}
\d \Gamma_\pi \dfn \d A_\pi - \frac{1}{2} \d C_{\pi \pi} \dfn \pi^* \d A - \frac{1}{2}  \pi^* (\d C) \pi
\end{equation}
is the growth differential, i.e., the instantaneous mean logarithmic growth rate, via use of the portfolio $\pi$. The \textbf{growth-optimal portfolio}
 is defined to maximise in a $(\omega, t)$-pointwise manner $\d \Gamma_\pi$, in the sense that  $\num = \arg \max_{\pi} \d \Gamma_\pi$. The solution to this quadratic maximisation problem satisfies the first-order conditions
\begin{equation} \label{eq:gop-unconstrained}
(\d C) \num = \d A.
\end{equation}

In order to avoid redundancies, we shall be making throughout the mild and natural assumption that $\d C$ has full rank\footnote{To be mathematically precise, we assume full rank in the $\big( \prob \otimes \sum_{i \in I} \int_0^\cdot (\d C_{ii} + |\d A_i|) \big)$-a.e.~sense.}, which means that the growth-optimal portfolio is given by
\begin{equation}\label{eq:gop}
	\num = (\d C)^{-1} \d A.	
\end{equation}
In that case, we have $\d \log X_\num = \d \Gamma_\num + \num^* \d M$, with maximal instantaneous growth
\begin{equation} \label{eq:210615}
\d \Gamma_\num = \frac{1}{2} (\d A)^* (\d C)^{-1} \d A,	
\end{equation}
and instantaneous squared investment volatility of
\begin{equation} \label{eq:gop_vola}
	(\d \log X_\num)^2 = \num^* (\d C) \num \equiv \d C_{\nu \nu} = (\d A)^* (\d C)^{-1} \d A = 2 \d \Gamma_\num.
\end{equation}
By \eqref{eq:gop}, \eqref{eq:210615}, and \eqref{eq:gop_vola}, $\d A_\nu = 2 \d \Gamma_\nu = \d C_{\nu \nu}$ holds, giving $2 \d \Gamma_\num = (\d A_\num)^2 / \d C_{\num\num}$.  In words, the instantaneous growth-rate of the growth-optimal portfolio equals half of its instantaneous squared Sharpe ratio.

We shall always assume that the nondecreasing integrated maximal growth process, given by

\[
	\Gamma_\nu \equiv \frac{1}{2} \int_0^\cdot (\d A)^* (\d C)^{-1} \d A,
\]
is real-valued, i.e., it does not explode.\footnote{In technical terms, this also ensures that both integrals in \eqref{eq:portfolio_growth}, when $\pi = \num,$ are well defined.} It is worthwhile noting that this existence of an integrable growth-optimal portfolio corresponds to a very weak no-arbitrage 
condition---see \cite{KK2007}.

\begin{remark}
	It should be stressed and recognised that growth comes at a high price for volatility. In a model with a level $\gamma$ of maximal (excess) growth, we need to endure  volatility of level $\sqrt{2 \gamma}$; see \eqref{eq:gop_vola}. As a matter of illustration, with $2 \%$ annual growth over risk-free investment, this already entails $20 \%$ annual volatility (which is more than what we tend to see in the market index). Along with the inherent difficulty of estimating rates of return, growth-optimal investment becomes quite tricky.	
\end{remark}

\section{Fund models} \label{S:FundModels}

This section treats the estimation of the growth-optimal portfolio in models with the property that the returns of certain funds span the returns of all assets, up to an orthogonal noise component. We first show that such models are characterised as exactly the ones where the growth-optimal portfolio  constrained to invest only in the funds is in fact globally growth-optimal. We then proceed in discussing frequentist non-parametric estimation of growth optimality. Section~\ref{S:Filtering} presents a ``Bayesian'' approach that eventually leads to robust shrinkage estimators in Section~\ref{S:Shrinkage}.

\subsection{Definition and characterisation of fund models}

Fix portfolio processes $(\funds^k; k \in K)$, where $K$ is a finite index set. These portfolios are to be understood as the only funds that individuals are able (or wish) to invest in, as opposed to having the freedom in investing in the whole universe of stocks. Typically, the number $|K|$ of funds will be much smaller than the number $|I|$ of available assets. Define also the matrix-valued processes 
\[
\funds \equiv (\funds_i^k; \, (i, k) \in I \times K).
\]
As above, and in order to facilitate reading, indices in $\Real^I$ will be subscripts, while indices in $K$ will be superscripts.

For matrix-valued processes $x \equiv (x_i^k; \, (i, k) \in I \times K)$ and $y \equiv (y_i^k; \, (i, k) \in I \times K)$ representing funds, we use the notation\footnote{We tacitly assume that any fund  $x \equiv (x_i^k; \, (i, k) \in I \times K)$ here and below constitutes a valid trading strategy; i.e., the integrals $\int_0^\cdot x^* (\d C) x$ and $\int_0^\cdot |x^* \d A|$ are finite valued.}
\[
\d C_{x y} \dfn  x^* (\d C) y
\]
for the $\Real^{K \times K}$-valued process of instantaneous covariations. To avoid unnecessary complications, we assume that $\d C_{\funds \funds}$ has full rank. Furthermore, we define
\[
\d A_x \dfn  x^* \d A
\]
for the $\Real^K$-valued process of instantaneous mean returns (or instantaneous risk premia) of the funds $x \equiv (x_i^k; \, (i, k) \in I \times K)$.

The growth-optimal portfolio when restricted to investment only in the funds represented by $\funds$ is $\funds \multi = \sum_{k \in K} \multi^k \funds^k$, where the process $\multi \equiv (\multi^k; \, k \in K)$ satisfies, in accordance to \eqref{eq:gop-unconstrained},
\begin{equation} \label{eq:gop-constrained}
(\d C_{\funds \funds}) \multi = \d A_\funds \quad \Longrightarrow \quad \multi =  (\d C_{\funds \funds})^{-1} \d A_\funds.
\end{equation}

A natural question is: when is the fund-constrained growth-optimal portfolio also growth-optimal for the whole market? In other words, when is the class of funds rich enough to already allow for growth-optimal investment? Here is a general result.

\begin{proposition} \label{prop:factor}
In the previously described market, the following two statements are equivalent:
\begin{enumerate}
	\item The (unconstrained) growth-optimal portfolio is such that $\num = \sum_{k \in K} \multi^k \funds^k$ for an appropriate process $\multi \equiv (\multi^k; \, k \in K)$, which then necessarily satisfies \eqref{eq:gop-constrained}.
	\item It holds that
	\begin{equation} \label{eq:factor_model}
	\d R_i = \sum_{k \in K} \beta_i^k  \d R_{\funds^k} + \d N_i, \qquad i \in I,
	\end{equation}
	for appropriate processes $\beta \equiv (\beta_i^k; \, (i, k) \in I \times K)$, where $(N_i; \, i \in I)$ are local martingales with the property that $(\d N_i)(\d R_{\funds^k}) = 0$ holds for all $k \in K$.
\end{enumerate}
\end{proposition}

Let us intuitively explain why enforcing the local martingale property on $(N_i; \ i \in I)$ ensures that the growth-optimal portfolio is a combination of the funds. Indeed, \eqref{eq:factor_model} implies that, investing in anything further than the funds, would just add extra volatility, without any risk compensation. A growth-optimal portfolio would never do this, as extra volatility would reduce growth.

Before giving the proof of Proposition \ref{prop:factor}, let us make an observation. There always exists a decomposition of the form \eqref{eq:factor_model} for appropriate processes $\beta \equiv (\beta_i^k;  (i, k) \in I \times K)$, where $(\d N_i)(\d R_{\funds^k}) = 0$ holds for all $k \in K$, as long as we do not insist that  $(N_i; \, i \in I)$ are local martingales. Indeed, we may identify $\beta$ via the orthogonality condition, noting that\footnote{Here, $(e_i; \, i \in I)$ are the usual basis unit vectors.} $e_i^* (\d C) \funds^{m} = \sum_{k \in K} \beta_i^k \d C_{\funds^k \funds^m}$ has to hold for all $i \in I$ and $m \in K$. We can then write $(\d C) \funds = \beta \d C_{\funds \funds}$, which gives $\beta = (\d C) \funds (\d C_{\funds \funds})^{-1}$.  With this definition of $\beta$, it is straightforward to check that $\d N_i \dfn \d R_i -  \sum_{k \in K} \beta_i^k  \d R_{\funds^k}$ is such that $(\d N_i)(\d R_{\funds^m}) = 0$ holds for all $m \in K$.

\begin{proof}[Proof of Proposition \ref{prop:factor}]
Assume that condition (1) is true, i.e., $\num = \funds \multi$, which gives $\d A = (\d C) \num = (\d C) \funds \multi$. Consider the decomposition \eqref{eq:factor_model} for $\beta \equiv (\beta_i^k; \, i \in I, \, k \in K)$ such that $\beta \d C_{\funds \funds} = (\d C) \funds$, so that $(\d N_i)(\d R_{\funds^k}) = 0$ holds for all $i \in I$ and $k \in K$, as explained right after the statement of Proposition \ref{prop:factor}. We only need to show that $(N_i; \, i \in I)$ is a family of local martingales. The previous relationships, together with \eqref{eq:gop-constrained}, give 
\[
\d A = (\d C) \funds \multi = \beta (\d C_{\funds \funds}) \multi = \beta \d A_{\funds}.
\]
Note then that the local drift differential of $\d N_i = \d R_i -  \sum_{k \in K} \beta_i^k  \d R_{\funds^k}$ equals
\[
\d A_i -  \sum_{k \in K} \beta_i^k \d A_{\funds^k} = \d A_i -  e_i^* \beta \d A_\funds = \d A_i -  e_i^* \d A = \d A_i - \d A_i = 0,
\]
ensuring that $N_i$ is a local martingale for all $i \in I$.

Conversely, assume that condition (2) holds true. Then, with $\beta$ satisfying $\beta \d C_{\funds \funds} = (\d C) \funds$, it holds that
\[
0 = \d A_i -  \sum_{k \in K} \beta_i^k \d A_\funds^k = \d A_i -  e_i^* \beta \d A_\funds  = e_i^* \left( 
\d A -  \beta \d A_\funds \right), \qquad i \in I,
\]
which gives 
\[
	\d A =  \beta \d A_\funds =  (\d C ) \funds (\d C_{\funds \funds} )^{-1} (\d A_\funds).
\]
 Setting $\multi \dfn (\d C_{\funds \funds})^{-1} (\d A_\funds)$, the previous reads $\d A =  (\d C) \funds \multi$, which implies that the growth-optimal portfolio $\num$ is such that $\num = \funds \multi = \sum_{k \in K} \multi^k \funds^k$.
\end{proof}

\begin{exa} \label{exa:CAPM}
	A special case of a single-fund model ($|K| = 1$) is the so-called \emph{Capital Asset Pricing Model} (CAPM) of \cite{sharpe1964capital} and \cite{Moss:66}, in which the single fund equals the \textbf{market portfolio} $\market \equiv (\market_i; \, i \in I)$, defined via
\begin{equation*}
	\market_i \dfn \frac{S_i}{\Sigma}, \quad i \in I, \qquad \text{where} \quad \Sigma := \sum_{j \in I} S_j.
\end{equation*}
Above, we give $(S_i; i \in I)$ the interpretation of market capitalisation\footnote{Note that this interpretation does not affect the relative dynamics in \eqref{eq:returns}.} (share price multiplied by number of shares outstanding); thus, $\Sigma$ becomes the total market capitalisation and $(\market_i; i \in I)$ the relative company capitalisations. With this notation, note that
\[
\frac{\d \Sigma}{\Sigma} = \sum_{j \in I} \frac{\d S_j}{\Sigma} = \sum_{j \in I} \market_j \frac{\d S_j}{S_j} = \d R_\market,
\]
which establishes that investing according to $w$ replicates the market capitalisation.

The CAPM states that $\d R_i = \beta_i \d R_\market + \d N_i$ holds for $i \in I$, for some processes $\beta \equiv (\beta_i; \, i \in I)$, where $(N_i; \, i \in I)$ are local martingales they are locally uncorrelated with the market, i.e., $(\d N_i)(\d R_\market) = 0$ holds for $i \in I$. This ``orthogonality'' gives
\[
\beta_i = \frac{\d C_{i \market}}{\d C_{\market \market}} = \frac{ e_i^* (\d C) \market}{\market^* (\d C) \market}, \quad i \in I.
\]

According to Proposition~\ref{prop:factor}, the CAPM is equivalent to the statement that the growth-optimal portfolio $\num$ equals $\multi \market$, where $\multi$ is a one-dimensional ``leverage'' process which would require estimation. In accordance to Merton's solution to the optimal investment problem, $\multi$ has the interpretation of ``local risk aversion'' for a representative agent in the market \citep{Merton1969,Merton1971}.

\end{exa}

\subsection{Estimation}  \label{SS:estimation}

Proposition \ref{prop:factor} implies that, for the purposes of growth-optimal investment, one need only estimate an unobservable $|K|$-dimensional process $\multi$, while utilising a potentially large amount of cross-sectional data across the returns of all the $|I|$ assets, providing hope in that estimation of the growth-optimal portfolio may be done more efficiently.

To simplify the exposition at this point, we take a purely frequentist and non-parametric point of view. We shall estimate locally in time excess returns, without further modelling assumptions. Although we are formally using differential notation, the understanding is that estimation is happening within a ``small'' window of observations. This makes the theory of estimation very general, but leads to very noisy estimates. As we shall argue below, only data from the fund returns are relevant for estimation of $\multi$; anything ``orthogonal'' to the process of fund returns is irrelevant; thus, unfortunately, cross-sectional data do not offer any advantage.

One may use the relationship $\d A = (\d C) \funds \multi$ to estimate $\multi$ by considering a combination $x \equiv (x_i^k; \, i \in I, k \in K)$ of funds. To this end, note that
\[
(\d C_{x \funds}) \multi = x^* (\d C) \funds \multi = x^* \d A = \d A_x \quad \Longrightarrow \quad \multi = (\d C_{x \funds})^{-1} \d A_x,  
\]
as long as $\d C_{x \funds}$ is of full rank. Therefore, we can use
\[
\widehat{\multi(x)} = (\d C_{x \funds})^{-1} \widehat{\d A_x}
\]
as an estimator, where $\widehat{\d A_x}$ is an unbiased estimator for $\d A_x$, with covariance matrix equal to $\d C_{x x}$.\footnote{Locally in time, $\d R_x$ is approximately multivariate Gaussian with mean vector $\d A_x$ and covariance matrix $\d C_{xx}$. In practice, we estimate $\d A_x$ locally estimate through the returns in a certain window of observations.} In continuous time, the instantaneous covariance matrix is observable and does not need to be estimated (for practical problems of estimating the covariance matrix with high-frequency data, see \citet{CMZ2020}).  The mean squared error in estimating $\multi$ is\footnote{The notation $\expec_{\cdot}$ for the expectations used here should be understood as conditional on the observations.}
\begin{align*}
\MSE (x) &= \expec_{\cdot} \left[ \norm{\widehat{\multi(x)} - \multi}^2 \right] = \expec_{\cdot} \left[  \norm{ (\d C_{x \funds})^{-1} \left(\widehat{\d A_x} - \d A_x\right)}^2 \right] \\
&= \trace  \left( (\d C_{x \funds})^{-1} (\d C_{xx}) (\d C_{\funds x})^{-1}\right).
\end{align*}
Here and in the sequel, ``$\trace$''  denotes the trace operator on matrices.

An application of Lemma \ref{lem:Frob_min} (with $\eta$ there being the identity matrix and $c$ there equalling $\d C$) implies that $\MSE (\funds) \leq \MSE (x)$ holds for all $x \in \Real^{I \times K}$ (such that $\d C_{x \funds}$ is non-singular). It follows that we may restrict attention in estimation through the fund $f$. Note that
\[
\MSE (\funds) = \trace \left( (\d C_{\funds \funds})^{-1} \right),
\]
which again shows that high fund volatility gives better estimation for $\multi$.

\begin{remark}\label{rem:linear_estimation_is_optimal}
	In such fully non-parametric setting, $\d R$ is formally conditionally multivariate Gaussian with mean vector $\d A = (\d C) \funds \multi$ and covariance matrix $\d C$. Therefore, in the class of unbiased estimators of $\theta$, the optimal one in terms of mean squared error will be linear in the data. It follows that $\widehat{\multi(f)}$ is (again, formally) the best linear unbiased estimator for $\multi$.
\end{remark}

Mean squared error is not the only objective that results in the fund $f$ being the most efficient way to estimate $\theta$ in the fund model, i.e., if the conditions of Proposition~\ref{prop:factor} are satisfied. Let us next consider minimising \textbf{distance from growth-optimality}, which is the drift differential $(\d \Gamma_\nu - \d \Gamma_{\widehat{\num}(x)})$ of $\d \log(X_{\num} / X_{\widehat{\num}(x)})$, with $\widehat{\num}(x) \dfn \funds \widehat{\multi} (x) = \funds (\d C_{x \funds})^{-1} \widehat{\d A_x}$. A standard application of It\^o's formula shows that, provided $\d C_{x \funds}$ is non-singular, this distance equals
\begin{align*}
	\frac{1}{2} \norm{ (\d C)^{1/2} \left(\widehat{\num}(x) - \num\right)}^2 &= \frac{1}{2} \norm{ (\d C)^{1/2} \funds (\d C_{x \funds})^{-1} \left(\widehat{\d A_x} - \d A_x\right)}^2 \\
	&= \frac{1}{2} \norm{ (\d C_{\funds \funds})^{1/2} (\d C_{x \funds})^{-1} \left(\widehat{\d A_x} - \d A_x\right)}^2.
\end{align*}
This distance corresponds to the utility loss due to estimation error in \citet{KS1996}, \citet{Avramov2004}, and \citet{KZ2007} for the case when the coefficient of relative risk aversion equals 1. 

We need to minimise 
\begin{align*}
	\DIS (x) &= \frac{1}{2} \expec_{\cdot} \left[ \norm{ (\d C_{\funds \funds})^{1/2} (\d C_{x \funds})^{-1} \left(\widehat{\d A_x} - \d A_x\right)}^2 \right] \\
	&= \frac{1}{2} \trace \left( (\d C_{\funds \funds})^{1/2} (\d C_{x \funds})^{-1} (\d C_{xx}) (\d C_{\funds x})^{-1} (\d C_{\funds \funds})^{1/2} \right).
	\end{align*}
 Lemma \ref{lem:Frob_min} again (with $\eta$ there equalling $(\d C_{\funds \funds})^{1/2}$) yields that  $\DIS (x)$ minimised at $x = \funds$. The minimal value equals
\[
\DIS (\funds) = \frac{1}{2} \trace(\id) = \frac{|K|}{2},
\]
which, very interestingly, \emph{does not depend on any market characteristic apart from the dimensionality of the fund}.

\section{Filtering and  estimation of growth}
\label{S:Filtering}

 The frequentist estimator of \S\ref{SS:estimation} is very noisy, due to the large error in the estimation of instantaneous returns. In order to have  practical and useful estimates, we move to a Bayesian formulation; here, even in fund models, other data (e.g., past and cross-sectional returns) may be important for estimation. For this purpose we consider a  filtering (learning) problem in this section. We study estimation problems related to growth-optimality in the presence of two different information flows. The finer ``theoretical'' information represents the one under which the model is specified, while the coarser ``practical'' information may represent an investor's lack of knowledge about (or observation of) some of the model parameters. While we work on a rather general and abstract setting, there is a specific illustration in  \S\ref{SS:Bayesian}, yielding a Bayesian estimate of the growth-optimal portfolio (in contrast to the previous ``frequentist'' estimates).

\subsection{Framework} \label{SS:FilteringFramework}

We work in a general two-level information model, for two information flows (filtrations) $\cF \equiv (\cF(t); t \geq 0)$ and $\cG \equiv (\cG(t); t \geq 0)$ with $\cF(t) \subseteq \cG(t)$ holding for all $t \geq 0$.

The investor's information flow corresponds to the information flow $\cF$. We assume that the returns process $R$ can be observed under the information flow $\cF$.\footnote{Without additional information,  $\cF$ would correspond exactly to the information flow generated by the return process $R$; however, we allow for setups where more information may be available to the investor.} In particular, this yields that the covariance process $C$ is observable under $\cF$.

The larger information flow $\cG$ is used here as a modelling device,  representing full information. In particular, as \S\ref{SS:Bayesian} below illustrates, such a framework includes the situation when the investor has a Bayesian prior on certain model parameters.

According to \eqref{eq:gop-unconstrained}, we write the decomposition of the returns $R \equiv (R_i; \, i \in I)$ under the larger information flow $\cG$ as
\begin{align*} 
\d R =  (\d C) \numg + \d M^\cG,	
\end{align*}
where $M^\cG$ is a continuous $\cG$-local martingale, $\d C = (\d R)(\d R)^*$, and $\numg$ is the growth-optimal portfolio under the information flow $\cG$.

We shall assume that the growth-optimal $\cG$-portfolio $\numg$ appearing in the above will be such that the nondecreasing process $G$, which is the integral of the differentials
\[
\d G \dfn \frac{1}{2}  \numg^*(\d C) \numg,	
\]
is well defined (i.e., finite). 
Note that $G$ is in fact the maximal achievable growth in the market under the finer information flow $\cG$.

	For a $\cG$-adapted stochastic process $\eta$, we shall use $\EF[ \eta ]$ to denote the process that at any point in time $t \geq 0$ equals the expected value $\expec[\eta(t)| \cF(t)]$; i.e., the best estimate of the value of $\eta(t)$ by an investor with information flow $\cF$.\footnote{In the jargon of the theory of stochastic processes, $\EF[ \eta ]$ is the $\cF$-optional projection of the process $\eta$.}
	
	Whenever $H$ is a finite variation process, we shall be slightly abusing notation and write $\EF [\d H ]$ to formally denote conditional expectation of the increments $\d H$ of $H$ given past information.\footnote{In mathematical terms, the integrated process $\int_0^\cdot \EF [\d H ]$ is the so-called dual $\cF$-optional projection of $H$. We will put assumptions in place so that this dual $\cF$-predictable projection exists when used.} In all the situations we shall encounter, there exist some continuous, nondecreasing, and $\cF$-adapted process $\clo$ such that $H = \int_0^\cdot \eta \d \clo$; then, we in fact have  $\EF [ \d H ] = \EF [\eta] \d \clo$.\footnote{This is a more precise definition of the dual $\cF$-optional projection of $H$. This expression does not depend on the processes $\eta$ and $\clo$, $\cG$-adapted and $\cF$-adapted, respectively, in the representation $\d H = \eta \d \clo$.}

We assume that
\[
\int_0^\cdot \EF [\numg^* (\d C) \numg] < \infty;
\]
i.e., that the optimal growth under information $\cG$ has a finite $\cF$--compensator. Since we are assuming that $\d C$ is of full rank, the above in particular implies that 
\[
\EF \big[ \norm{\numg}^2 \big]	< \infty.
\]
This allows us to define the processes $\numf$ and $\varnu$, respectively, as being the mean and covariance matrix of the $\cF$-conditional law of $\numg$: 
\[
	\numf \dfn \EF [\numg]; \qquad \varnu \dfn  \EF [(\numg - \numf) (\numg - \numf)^*] = \EF [\numg \numg^*] - \numf \numf^*.	
\]

In the finer information, the growth-optimal portfolio is given by the process $\numg$. An investor equipped with only the coarser information might not be able to observe $\numg$, which creates uncertainty on the values of the process $\numg$. Such investors' best guess provided the information flow $\cF$ is $\numf$, obtained by ``filtering'' the data. Moreover, conditionally on the coarser information, $\numg$ has (conditional) second moments described by the matrix-valued process $\kappa$. Note that we do not enforce any structure on the (conditional) law of $\numg$, apart from the existence of conditional second moments, which comes as a consequence of absence of arbitrage in the finer information through the fact that $G$ is a finite-valued process.

\subsection{Loss of growth due to filtering}

The drift in the dynamics of $R$ for the investor under the information flow $\cF$ is ``filtered'' from the corresponding drift of the dynamics under $\cG$. With the above assumptions and notation, the dynamics of $R$ under $\cF$ become  $\d R = (\d C) \numf + \d M^\cF$, where $M^\cF$ is an $\cF$-local martingale. These dynamics yield directly that  $\numf$ is the growth-optimal portfolio under information flow $\cF$. The maximal achievable $\cF$-growth $F$ hence satisfies
\begin{align} \label{eq:200722}
\d F \dfn \frac{1}{2}  \numf^* (\d C) \numf.	
\end{align}

Consider now a portfolio $\pi$ from the side of the investor, that uses only the coarser information flow $\cF$. Then, according to \eqref{eq:portfolio_growth}, its local growth measured under the information flows $\cF$ and $\cG$, respectively, is given by
\begin{align} \label{eq:200627.2}
\d \Gamma^{\cF}_\pi = \pi^* (\d C) \numf - \frac{1}{2} \pi^* (\d C) \pi; \qquad \d \Gamma^{\cG}_\pi = \pi^* (\d C) \numg - \frac{1}{2} \pi^* (\d C) \pi.	
\end{align}
In particular, $\EF [\d \Gamma_\pi^{\cG}] = \d \Gamma_\pi^{\cF}$, and
\begin{align} \label{eq:200627.3}
	\EF \left[\left(\d \Gamma_\pi^{\cG} - \d \Gamma_\pi^{\cF} \right)^2\right]	= \norm{\varnu^{1/2} (\d C) \pi}^2.
\end{align}
It follows that the best estimate for the portfolio's local $\cG$-growth coincides with its $\cF$-growth, and there is a closed-form expression for the $\cF$-conditional variance of the local $\cG$-growth.

However, when we want to compare the maximal local growth $\d F$ and $\d G$ between the two information flows $\cF$ and $\cG$, respectively, the previous does not apply, because the $\cG$-growth-optimal portfolio $\numg$ cannot be observed under the coarser information flow $\cF$. In fact, we have
\begin{align}
	2 \EF [\d G] =  \EF [\numg^* (\d C) \numg] &= \numf^* (\d C) \numf + 2 \numf^* (\d C) \EF [\numg - \numf] + \EF [(\numg - \numf)^* (\d C) (\numg - \numf)]  \nonumber\\
	&= \numf^* (\d C) \numf + \trace (\varnu^{1/2} (\d C) \varnu^{1/2}) \nonumber\\
		&= 2 \d F + \trace (\varnu \d C). 	  \label{eq:200126.2}
\end{align}
The term $\EF [\d G] - \d F =  (1/2)\trace (\varnu \d C) \geq 0$ measures the ``distance'' of the maximal $\cF$-growth differential from the expected maximal $\cG$-growth differential, when seen under information $\cF$. This quantity $\EF [\d G] - \d F$ can also be regarded as loss of growth differential coming from the estimation procedure of $\numg$ by $\numf$.

\subsection{Loss of growth increases as the investment universe does} \label{SS:loss}

Clearly, restrictions in investment will lead to loss of growth in both informational levels. What is also true is that \emph{loss of growth} when only using information flow $\cF$, as described above, also decreases once the investment universe that an investor tries to utilise becomes smaller. This is intuitively reasonable, as the corresponding estimation problem becomes easier.

Formally, recalling the setting of Section~\ref{S:FundModels}, assume that one may only trade in $\cF$-adapted funds 
\[
\funds \equiv (\funds_i^k; \, (i, k) \in I \times K),
\]
representing available investment opportunities. Assuming throughout that $\d C_{\funds \funds}$ is invertible, the constrained $\cG$-growth-optimal portfolio $\num_\funds$ is then given by
\[
\num_\funds := f (\d C_{\funds \funds})^{-1} \funds^* \d A = f (\d C_{\funds \funds})^{-1} \funds^* (\d C) \num.
\]
The maximal achievable $\cG$-growth differential under such restriction on fund investment equals
\begin{align*}
	\d G_\funds &= \num^* (\d C) \funds (\d C_{\funds \funds})^{-1} f^* \d A - 
	\frac{1}{2}  \num^* (\d C) \funds (\d C_{\funds \funds})^{-1} f^* (\d C)  f (\d C_{\funds \funds})^{-1} \funds^* (\d C) \num\\
		&= \frac{1}{2}  \num^* (\d C) \funds (\d C_{\funds \funds})^{-1} f^* (\d C)  \num.
\end{align*}
Similarly, for the maximal achievable $\cF$-growth differential in this investment universe we have
\begin{align*}
	\d F_\funds &=  \frac{1}{2}  \numf^* (\d C) \funds (\d C_{\funds \funds})^{-1} f^* (\d C)  \numf.
\end{align*}
As in \eqref{eq:200126.2}, the ``distance'' of the maximal $\cF$-growth differential from the expected maximal $\cG$-growth differential under this restricted investment satisfies
\begin{align} \label{eq:201121}
	\EF [\d G_\funds]  - \d F_\funds = \frac{1}{2}  \trace \left(\kappa (\d C) \funds (\d C_{\funds \funds})^{-1} f^* \d C \right) \leq  \frac{1}{2}  \trace \left(\kappa \d C\right) = \EF [\d G]  - \d F,
\end{align}
where the above inequality follows from  Lemma~\ref{lem:error_reduction}.

The same argument shows that whenever another fund $\bar \funds$, of potentially different dimensionality $I \times \bar K$, represents further restrictions from $f$ (in the sense that there is an appropriate $\cF$-adapted matrix-valued process that maps $f$ to $\bar f$), then
\[
	\EF [\d G_{\bar \funds}] - \d F_{\bar \funds} \leq \EF [\d G_\funds] - \d F_\funds.
\]
In words, the less investment opportunities are available, the less is the loss in growth coming from the estimation of the growth-optimal portfolio.

\begin{remark} \label{R:200627}
	As in Section~\ref{S:FundModels} we could assume that returns follow a certain fund structure under $\cG$, relying on the $\cF$-adapted funds $\funds \equiv (\funds_i^k; \, (i, k) \in I \times K)$; i.e., that \eqref{eq:factor_model} holds for appropriate   processes $(\beta_i^k; \, i \in I, k \in K)$, $\cF$-adapted  funds $\funds$, and $\cG$-local martingales $(N^\cG_i; \, i \in I)$ such that $(\d N^\cG_i)(\d R_{\funds^k}) = 0$ for all $i \in I$ and $k \in K$. Thanks to the paragraph following the statement of Proposition~\ref{prop:factor}, the processes $(\beta_i^k; \, i \in I, k \in K)$ may be chosen $\cF$-adapted (i.e., observable under the coarser information). Using the appropriate filter, one sees that the fund structure also holds under $\cF$, of course with different $\cF$-local martingales $(N^\cF_i; \, i \in I)$ such that $(\d N^\cF_i)(\d R_{\funds^k}) = 0$ for all $i \in I$ and $k \in K$.

	With the notation setup previously, and thanks to Proposition~\ref{prop:factor}, we have $\d G_f = \d G$ and $\d F_f = \d F$. Under the assumed factor structure on the returns, the (unconstrained) growth-optimal portfolio only invests in the funds under both information structures $\cF$ and $\cG$.  As a direct implication, the inequality in \eqref{eq:201121} is actually an equality.

	Consider now two models I and II for the returns of a given family of stocks, and assume that the two models agree on the dynamics $R_\funds \dfn \funds^* \d R$ of the returns of a fixed set of $\cF$-adapted funds $\funds$. Suppose also that model~I satisfies the fund structure of Section~\ref{S:FundModels}, whereas model~II does not.\footnote{One way to regard model~II is a situation where the investor does not have correct information about the factor structure \citep[see, e.g.,][]{FGX2020, GLX2021, GX}.} The previous considerations imply that there is larger loss in optimal growth passing from the $\cG$ to the $\cF$ information flow in the global investment universe under model~II than under model~I. (Indeed, in both models the growth loss in the performance of investments in the funds only are the same; in the fund model~I this actually corresponds to the total growth loss, whereas in the non-fund model~II the growth loss will be greater if we consider the full investment universe.) 
	
Assume now that model~I represents the true data-generating mechanism but the econometrician uses the misspecified model~II	instead.  Since the econometrician estimates the wrong quantity,  less growth will be obtained. 
Therefore, it may be the case that the econometrician is under the wrong impression that lower growth comes from large \emph{estimation} error, rather than large \emph{model misspecification} error.	
\end{remark}

\subsection{An example: Bayesian updating} \label{SS:Bayesian}

\subsubsection{Growth-optimal portfolio with Gaussian prior}

Under a probability measure $\qprob$ (which will be \emph{risk-neutral} in our model), let $R$ be a continuous $\cF$-local martingale with $R(0)$ to be defined later on. Let $\numg$ be an independent from $\cF$ random variable having Gaussian law with given mean $\numf(0) \in \bbR^I$ and covariance matrix $\varnu(0) \in \bbR^{I \times I}$. Let $\cG$ be the smallest information flow containing $\cF$ and knowledge of $\numg$ from the beginning of time. We set $\d C := (\d R)(\d R^*)$, and set $C(0) := \varnu(0)^{-1}$ and $R(0) := \varnu(0)^{-1} \numf(0)$. These values of $C(0)$ and $R(0)$ do not affect the differentials $\d R$ and $\d C$, and as a consequence will not affect trading; however, they are convenient in writing easier formulae, and are also very interpretable in the empirical Bayesian setup of \S\ref{sss:empiricalBayes}.

Continuing, define the probability $\prob$ such that
\[
\frac{\d \prob}{\d \qprob} \Big|_{\mathcal{G}(\cdot)} = \exp \left(  \int_0^\cdot \numg^* \d R - \frac{1}{2} \int_0^\cdot \numg^* (\d C) \numg \right),
\]
with the underlying assumption that the process above on the right-hand-side is a true (and not just local) $(\cG, \qprob)$-martingale. We shall be working under this probability $\prob$. Note that the $\cG$-drift differential of $R$ under $\prob$ equals $(\d C) \numg$.

\begin{lemma} \label{L:200727}
	The $\cF$-conditional law of $\numg$ is Gaussian with mean vector $\numf = C^{-1} R$ and covariance matrix $\varnu= C^{-1}$.	
\end{lemma}

\begin{proof}
	To see this, fix some Borel set $A$ and note that
	\begin{align*}
		\log \PF [\numg \in A]\, & \sim\, \log \EF_\qprob\left[\mathbf{1}_{\{\numg \in A\}} 
			\exp \left(  \int_0^\cdot \numg^* \d R - \frac{1}{2} \int_0^\cdot \numg^* (\d C) \numg \right) \right] \\
			&\sim\, \log \int_A
			\exp \left(  x^* (R - R(0)) - \frac{1}{2}  x^* (C-C(0)) x + \ell (x) \right)  \d x,
			\end{align*}
	where $\ell(x) = x^* R(0) - (1/2) x^* C(0) x$ is (up to an additive constant) the prior log-density of $\numg$. Here ``$\sim$''  denotes equality up to an additive normalising process that does not depend on the Borel set $A$. Therefore, we obtain that the $\cF$-conditional log-density of $\numg$ equals (up to an additive normalising process)
	\[
	\log \frac{\PF [ \numg \in \d x]}{\d x} \sim \ell(x) + x^* (R - R(0)) - \frac{1}{2} x^* (C - C(0)) x =  x^*R - \frac{1}{2} x^* C x.
	\] 
	The result is now immediate.
\end{proof}

In this specific case, where $\varnu = C^{-1}$, we obtain from \eqref{eq:200126.2} that
\begin{align} \label{eq:200627}
\EF [\d G] - \d F = \frac{1}{2} \trace \left( C^{-1} \d C \right)	= \frac{1}{2} \d \log (\det (C)),
\end{align}
where ``$\det$'' denotes matrix determinant.

\subsubsection{Empirical Bayes setting}\label{sss:empiricalBayes}

In the empirical Bayesian setup, one uses past data to estimate the prior law. Assume that there already have been  $\delta > 0$ previous years of observation, in which case we set $C(- \delta) = 0$ and $R (- \delta) = 0$, and count everything from time $- \delta$ onward. At time $- \delta$, we interpret as ``uninformative'' prior a sequence of Gaussian laws with some fixed mean and covariance matrix that ``explodes'' in the sense that its inverse converges to zero. Then, it is easily seen by a limiting argument that the posterior law at time zero of $\numg$ is Gaussian with mean $\numf(0) = C(0)^{-1} R(0)$ and covariance matrix $\varnu(0) = C(0)^{-1}$. This is exactly the reason for ``reverse engineering'' the values of $C(0)$ and $R(0)$ to be consistent with these formulae. In fact, it should be noted that one does not need Gaussian prior laws at time $-\delta$ for this to hold---any sequence of prior laws with the property that conditional laws on bounded intervals are asymptotically uniform will lead to the same Gaussian limit. 
	
Continuing in the empirical Bayesian setup, assume further that, the local volatility matrix of the returns is constant with respect to an operational clock, the latter modelled via a nondecreasing $\Real$-valued, continuous, and adapted process $\clo$. One could interpret $\clo$ as market activity (as opposed to calendar) time, and the assumption of constant volatility with respect to it means that $\d C = c \d \clo$ and $C(0) = c \clo(0)$ holds for a constant positive definite matrix $c$. We note that $\clo(0) > 0$, in order to model previous observations, in accordance to \S\ref{sss:empiricalBayes} above. In this special case, we then obtain from \eqref{eq:200627} that
\[
\expec^{\cF} [\d G] - \d F = \frac{1}{2} \trace \left( C^{-1}  c \d O \right) = \frac{|I|}{2}  \frac{\d O}{O} = \frac{|I|}{2} \d \log (O).
\]
Interestingly, and as was noted in the analysis of Section \ref{S:FundModels}, the distance from optimality depends only on the dimensionality of the assets (and here, also on the previous years of observation), but \emph{not} on any market characteristic. 

To get a feeling for the numbers, with only a single fund (recall Remark~\ref{R:200627}) and $\clo(0) = 10$ (for example, this could represent 10 calendar years of observation if $\clo$ is calendar time), one still loses at the beginning of trading instantaneously a vast $(2 \clo(0))^{-1} |I| = .05 = 5 \%$ of growth (in absolute terms, and with respect to the operational clock change $\d O$).

Of course, the fact that there is more average distance when the number of the funds is higher does not need the assumption of constant covariance matrix; as the discussion before Remark~\ref{R:200627} implies, it is always true.

\subsubsection{Trunctated Gaussian priors}\label{sss:restricted_Gaussian}

We have considered here Gaussian priors on the growth-optimal $\numg$. Via a straightforward adjustment, we may accommodate situations when we  have further prior information on the possible range of values that $\numg$ may take. To this end, let us consider some set $U \subset \Real^{I}$ with positive Lebesgue measure. We assume that the prior law of $\nu$ is  $\indicator{U} \exp(\ell)$, where  $\ell(x) \sim x^*R(0)x - (1/2) x^* C(0) x$ (with ``$\sim$'' denoting equality up to an additive normalising constant), as in the proof of Lemma~\ref{L:200727}; that is, the prior on $\nu$ corresponds to a normal distribution with mean vector $C(0)^{-1} R(0)$ and covariance matrix $C(0)^{-1}$, conditioned on taking values in $U$. It then follows as in Lemma~\ref{L:200727} that the $\cF$-conditional law of $\numg$ is and Gaussian with mean $\numf =  C^{-1} R$ and covariance matrix $\varnu = C^{-1}$, conditioned on taking values in $U$. It follows that
\begin{align*}
	\numf &\dfn \EF [\numg] = \frac{\int_U x \exp(x^* R  - (1/2) x^* C x) \d x}{\int_U \exp(x^* R  - (1/2) x^* C x) \d x}; \\
	\varnu  &\dfn   \EF [(\numg - \numf) (\numg - \numf)^*] = 
	\frac{\int_U x x^* \exp(x^* R  - (1/2) x^* C x) \d x}{\int_U \exp(x^* R  - (1/2) x^* C x) \d x} - \numf \numf^*.
\end{align*}

Particular closed-form expressions can be retrieved for the one-fund case (i.e., $|I| = 1$), when $U = (l, r)$ for some $- \infty \leq l < r \leq \infty$. (For example, one may want the prior to be supported only on the nonnegative real line). 
Under these assumptions,  we obtain that the $\cF$-conditional law of $\numg$ is truncated normal with 
\begin{align*}
	\numf &= \frac{R}{C} + \frac{1}{\sqrt{C}}  \frac{-\phi(\bar r) + \phi(\bar l)}{\Phi(\bar r) - \Phi(\bar l)}; \\
	\varnu &= 
	\frac{1}{{C}} \left(1 + \frac{- \bar r \phi(\bar r) + \bar l \phi(\bar l) }{\Phi(\bar r) - \Phi(\bar l)}
	- \left(\frac{\phi(\bar r) - \phi(\bar l)}{\Phi(\bar r) - \Phi(\bar l)}\right)^2
	\right),
\end{align*}
where $\bar l = l \sqrt{C} - R/\sqrt{C}$, $\bar r = r \sqrt{C} - R/\sqrt{C}$, and $\phi$ (respectively, $\Phi)$ denotes the probability density (respectively, cumulative distribution) function of a standard normal law.

\section{Shrinkage} \label{S:Shrinkage}

In this section we study an interesting ``shrinkage'' estimator of the growth-optimal portfolio. When filtering from $\cG$ down to information $\cF$ in the context of Section \ref{S:Filtering}, one obtains an estimate $\numf$ for the $\cF$-growth-optimal portfolio that maximises the $\cF$-expectation of the growth differential $\d \Gamma^\cG_\pi$ over all $\cF$-portfolios $\pi$. This is certainly an aggressive strategy leading to maximal growth $\d F = \EF [\d G]$; however, from the vantage point of the $\cF$-investor there may be a lot of ``spread'' in the $\cF$-conditional law of $\d \Gamma^\cG_\numf - \d F$. It may be more appealing to take a slightly more conservative approach, and instead try to minimise the spread of $\d \Gamma^\cG_\pi - \d F$ over $\cF$-portfolios $\pi$, thereby targetting maximal growth with the least amount of deviation, even if not fully achieving such maximal growth in expectation. This approach allows for the introduction of some ``bias'' in the estimation of the growth-optimal portfolio in order  to reduce the overall ``variance'' of the  target, namely the growth differential. As we shall see in this section, this results in ``shrinkage'' of the portfolio $\numf$, which has the additional welcome effect that overall volatility of the resulting wealth is reduced. This way, two reductions occur simultaneously. On the one hand, distance from optimal growth with respect to uncertainty in filtering is reduced, which only affects the drift of the log-wealth process. On the other hand, intertemporal reduction of overall portfolio volatility is achieved, which \emph{a priori} has nothing to do with growth.

\subsection{General framework} \label{SS:Shrinkage_General}

We keep the setup and notation from Section~\ref{S:Filtering}. 

The actual instantaneous $\cG$-growth of the portfolio $\numf$ equals
\[
\d \Gamma^\cG_{\numf} = \numf^* \d A - \frac{1}{2} \numf^* (\d C) \numf = \numf^* (\d C) \numg - \frac{1}{2} \numf^* (\d C) \numf.	
\]
We wish to find an $\cF$--measurable portfolio $\pi$ whose $\cG$-growth is as close as possible to the optimal estimated $\cF$-growth, but $\numf$ may not be the best way to achieve this.

Using an $\cF$-predictable portfolio $\pi$, \eqref{eq:200627.2} gives that its $\cG$-growth satisfies
\[
\d \Gamma^\cG_\pi - \d F = - \frac{1}{2} \norm{(\d C)^{1/2} (\pi - \numf)}^2 + \pi^* (\d C) (\numg - \numf) .
\]
Hence one can write 
\begin{align} \label{eq:200628.2}
	(\d E_\pi)^2 = \EF \left[(\d \Gamma^\cG_\pi - \d F)^2 \right] =  \frac{1}{4} \norm{(\d C)^{1/2} (\pi - \numf)}^4 + \norm{\varnu^{1/2} (\d C) \pi}^2
\end{align}
for the average ``distance'' in the sense of mean squared error of the $\cG$-growth of $\pi$ from the optimal $\cF$-growth. On the right hand side of \eqref{eq:200628.2}, note the appearance of the variance term of \eqref{eq:200627.3}, as well as the squared bias term $\norm{(\d C)^{1/2} (\pi - \numf)}^4$.

We wish to (pointwise) minimise $\d E_\pi$ over portfolios; i.e., to solve
\[
\rho \dfn \arg \min_{\pi} \d E_\pi,
\]
and then define $\d E \dfn \d E_\shrunk$. An invocation of Lemma \ref{lem:shrink_multi_d} (with $z$, $y$, and $h$ there equal to $(\d C)^{1/2} \numf$, $(\d C)^{1/2} \pi$, and $(\d C)^{1/2} \varnu (\d C)^{1/2}$, respectively) and straightforward algebra give  
\begin{align} \label{eq:shrunk_portfolio}
	\shrunk = \left( \id + \varnu \frac{\d C}{\d B}  \right)^{-1} \numf, 
\end{align}
where\footnote{In fact, it holds that $\d B > 0$, except in the degenerate case where $\kappa (\d C) \numf = 0$, in which case $\rho = \numf$. The precise statement is given in Lemma \ref{lem:shrink_multi_d}.}
$0 \leq \d B \leq (1/2) \d C_{\numf \numf} = \d F$,  and $\d B$ solves
\begin{align} \label{eq:200628.4}
	\frac{1}{2} \norm{ \left( (\d C)^{1/2} \varnu (\d C)^{1/2} + (\d B) \id \right)^{-1} (\d C)^{1/2} \varnu (\d C) \numf }^2 = \d B.	
\end{align} 
From the proof of Lemma \ref{lem:shrink_multi_d} and from \eqref{eq:200627.2}, the increment
\[
\d B = \frac{1}{2} \norm{(\d C)^{1/2} (\shrunk - \numf)}^2 = \d F - \d \Gamma^\cF_\shrunk = \d \Gamma^\cF_\numf - \d \Gamma^\cF_\shrunk
\]
gives the local reduction in $\cF$-growth of the portfolio $\shrunk$ with respect to $\numf$. 
Furthermore, Remark \ref{rem:on_shrink_multi_d} gives
\begin{align*}
	\d C_{\numf \numf} = \norm{(\d C)^{1/2} \numf}^2 &= \norm{(\d C)^{1/2} \shrunk}^2 + 2 \frac{(1/4) \norm{(\d C)^{1/2} (\shrunk - \numf)}^4 + \norm{\varnu^{1/2} (\d C) \shrunk}^2}{(1/2)\norm{(\d C)^{1/2} (\shrunk - \numf)}^2} \\
	&= \d C_{\shrunk \shrunk} + 2 \frac{(\d E)^2}{\d B}.
\end{align*}
The last equality shows how the reduction of the local variance that portfolio $\shrunk$ achieves with respect to $\numf$ is connected to the optimal squared distance $(\d E)^2$ in following the $\cF$-growth $\d F$, as well as the resulting loss $\d B$ of  $\cF$-growth.

\subsection{Uniform shrinkage}
\label{SS:UniformShrink}

In the previous subsection we have characterised the optimal portfolio whose growth minimises the average distance to the optimal $\cF$-growth. The expression in \eqref{eq:shrunk_portfolio} is not explicit, as one needs to determine $\d B$. A fast and efficient numerical algorithm is suggested in Remark~\ref{rem:on_fixed_point}. However, one does need for input an estimate for $\d C$. Such estimate can be quite difficult to obtain in an efficient way, even with relatively high-frequency data. We will see below that in certain cases this can be avoided.

A more explicit expression from the one in \eqref{eq:shrunk_portfolio} may be obtained if one minimises \eqref{eq:200628.2} only over portfolios of the form $a \numf$ for some process $a$; i.e., one shrinks all positions in the $\cF$-growth-optimal portfolio uniformly.  Indeed,  \eqref{eq:200628.2} simplifies then to
\begin{align*}
	(\d E_{a \numf})^2 =  (1-a)^4 (\d F)^2 + a^2 \norm{\varnu^{1/2} (\d C) \numf}^2
\end{align*}
Defining $\d V \dfn \norm{\varnu^{1/2} (\d C) \numf}$, we minimise $\Real \ni a \mapsto \xi (a) \dfn (\d F / \d V)^2 (1 - a)^4 + a^2$ over $a \in \Real$.

There is in fact a unique minimiser for this problem, and it is $[0,1]$-valued. To wit, note that $\xi'(a) = - 4 (\d F / \d V)^2 (1 - a)^3 + 2 a$, which is increasing in $a$ (differentiate once again and note that the derivative is positive) and such that $\xi'(0) = - 4 (\d F / \d V)^2 \leq 0 \leq 2 = \xi'(1)$. The solution can be given in closed form (using Cardano's formula and some algebra) as
\begin{align} \label{eq:200628.3}
a = 1 - \frac{3}{1 + (\sqrt{1+\psi} + \sqrt{\psi})^{2/3} + (\sqrt{1+\psi} + \sqrt{\psi})^{-2/3}}, \qquad \text{where} \quad \psi = \frac{27}{2} \left( \frac{\d F}{\d V} \right)^2.
\end{align}

In the one-fund case, ``uniform'' shrinkage actually provides optimal shrinkage as in \S\ref{SS:Shrinkage_General}, and we have
\[
\psi = \frac{27}{8} \frac{\numf^4 (\d C)^2}{ \varnu (\d C)^2 \numf^2  } = \left( \frac{3}{2} \right)^3 \frac{\numf^2}{\varnu}.	
\]
It is important to note that in this one-fund case, $\psi$ will not depend on the estimation of $\d C$. Furthermore, in the Bayesian setup of \S\ref{SS:Bayesian}, where we have $\numf^2 / \kappa = R^2 / C$, the closed-form expression for $\psi$ (and, therefore, for $a$ as well) only involves integrated quantities, which are far more stable and robust to estimate.

As in the discussion of \S\ref{sss:empiricalBayes} in the empirical Bayes framework, let us now consider an arbitrary number of funds, but with $\d C = c \d \clo$ and $C(0) = c \clo(0)$ for a constant matrix $c$ and a nondecreasing $\Real$-valued, continuous, and $\cF$-adapted operational clock process $\clo$. Upon using the facts that $\kappa = C^{-1} = \clo^{-1} c^{-1} $ and $\numf = C^{-1} R = \clo^{-1} c^{-1} R$, the representation in \eqref{eq:shrunk_portfolio} yields, after straightforward but slightly tedious computations, that
\[
\rho = \frac{\d B}{\d B + \d \log \clo} \numf;	
\]
i.e., we have again a uniform shrinkage parameter $a = \d B / (\d B + \d \log \clo) \in [0,1]$.  By the previous computations, $a$ satisfies \eqref{eq:200628.3}, now with
\[
\psi = \left( \frac{3}{2} \right)^3  \norm{c^{1/2} \numf }^2 O = \left( \frac{3}{2} \right)^3  R^* C^{-1} R.	
\]
Remark~\ref{rem:reduction_multi_d} provides a check for these computations by directly relating equation \eqref{eq:200628.4} for $\d B$ with the cubic equation $\xi'(a) = 0$  for $a$ above.  

To conclude,  in both of these cases, namely in the one-fund case and in the case of constant covariance rate (with respect to the operational clock $\clo$) and Bayesian prior,  uniform shrinkage is optimal: the optimal shrinkage portfolio $\rho$ of \ref{SS:Shrinkage_General} is given by an explicit constant multiplier of $\numf$.

\section{Empirical study}  \label{S:empirics}

In this section we show how the shrinkage estimators perform. We consider the one-fund case with the market portfolio assumed to be the single fund (see Example \ref{exa:CAPM}). Here we only show the results for the US market, for which we have the longest time series.  We use the \emph{Value-Weighted Return index with dividends}, obtained from the  Center for Research in Security Prices (CRSP), and the one-month treasury bill rates as risk-free rates, obtained from the Fama-French dataset. The CRSP time series is available from January 1927, and the Fama-French time series from July 1927.  Hence, we start the empirical study July 1927, and run it until December 2020.

The online appendix (currently Appendix~\ref{A:data})  provides the same empirical analysis as done here for the US market for four more equity markets (UK, Germany,  Italy,  Australia). The time series for these four markets are shorter; hence the results less convincing.

We have daily returns that are adjusted for dividends. These returns are then turned into excess returns by adjusting them with the corresponding risk-free rate. To obtain the integrated return process $R$ we sum up these daily returns. To obtain the integrated variance process $C$ we sum up the squared returns.

\begin{figure}
	\includegraphics[width=1\textwidth]{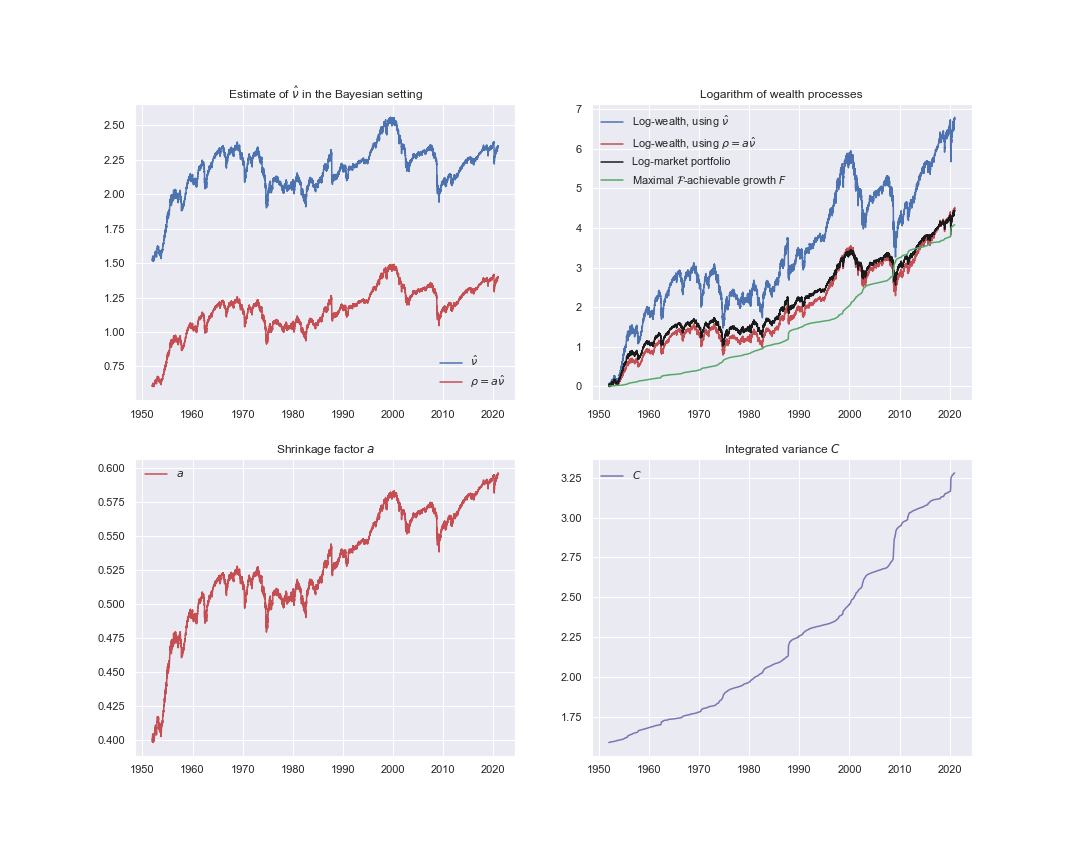}
	\caption{US data. See the main text for further explanations.}
	\label{F:US}
\end{figure}

 We present the results in Figure~\ref{F:US} in four panels.  Each panel describes a time series. We begin these time series 7500 trading days (i.e., about 30 years) after the first data are available.  In the notation of \S\ref{sss:empiricalBayes}, this corresponds to $\delta \approx 30$. 
 
The upper left panels show the estimate of the growth-optimal portfolio $\numf$ in the Bayesian setup of \S\ref{SS:Bayesian} (recall  Lemma~\ref{L:200727}) and of the shrunk portfolio $a \numf$. Here the optimal shrinkage factor $a$ is computed by \eqref{eq:200628.3}. The lower left panels display $a$.  The upper right panel displays several wealth processes (in logarithmic units) and the maximal $\cF$-achievable growth $F$, described in \eqref{eq:200722}.  The wealth processes correspond to the market portfolio,  the growth-optimal portfolio $\numf$ and its shrank version, $a\numf$. All three processes are standardised to be one at the first displayed date. Since excess returns are used, these wealth processes are discounted indeed discounted wealth processes, i.e., show the outperformance over a portfolio that holds the risk-free asset only.  The lower right panel displays the quadratic variation process $C$, approximated by the cumulative sum of squared fund returns.

We can see that the shrinkage term $a$ tends to be much lower than $1$. This is true for the US data, for which we have over ninety years of observations, but $a \approx 0.6$.  In the other four countries, $a$ tends to be less than $0.5$ (UK) or $0.3$ (Germany, Italy, Australia);  see  Appendix~\ref{A:data}.  Shrinkage leads to a clear reduction of volatility. The logarithmic wealth process corresponding to the shrunk portfolio $a \numf$ also tends to track the maximal $\cF$-achievable growth $F$ better than the logarithmic wealth process corresponding to the $\numf$ does. 

We also performed the same study with a restricted Gaussian prior, as suggested in \S\ref{sss:restricted_Gaussian}. None of the results changed significantly; hence we omit the corresponding panels here.

\section{Conclusion} \label{S:conclusion}

We have studied estimation of the growth-optimal portfolio and derived efficient estimates under the  fund models. We have also investigated estimation and filtering in a two-informational setting. We have shown, in particular, that the expected loss of the growth rate is larger, the larger the investment universe. We have proposed a shrinkage method, targeting maximal growth with the least amount of deviation from the optimal growth rates.

We have conducted an empirical analysis  under the assumption of a one-fund model (CAPM) and shown that the estimated portfolio with the shrinkage method exhibits stable returns without losing much of the growth rate.  It would be interesting to extend the analysis based on  multi-fund models similar to those proposed by \citet{KNS2020}, \citet{GKX2020}, and \citet{GX}, and to investigate the performance of the shrinkage portfolio in comparison with the unrestricted  growth optimal portfolio.

\appendix

\section{Technical Results}

\subsection{An auxiliary lemma for \S\ref{SS:estimation}}

We provide here a technical lemma in linear algebra, which is used in Section~\ref{S:FundModels}. To this end, 
 we let $\norm{\cdot}_F$ denote the Frobenius matrix norm in the space $\Real^{K \times K}$; i.e., $\norm{x}_F^2 \dfn \inner{x}{x}_F$, for the inner product $\inner{x}{y}_F \dfn \trace(x^* y)$ for $x, y \in \Real^{K \times K}$.   Moreover, in analogy to Section~\ref{S:FundModels}, for  $c \in  \Real^{I\times I}$ and  $x,y \in  \Real^{I\times K}$ we write $c_{xy} := x^* c y$.

\begin{lemma} \label{lem:Frob_min}
	For an arbitrary symmetric positive-definite $c \in \Real^{I \times I}$,  and arbitrary $\eta \in \Real^{K \times K}$, $\funds \in \Real^{I \times K}$,
	 the function
	\[
	\Real^{I \times K} \ni x \mapsto \norm{c_{xx}^{1/2} c_{\funds x}^{-1} \eta}^2_F,
	\]	
	(with the understanding that it has finite value only for $x \in \Real^{I \times K}$ such that $c_{x \funds}$ is non-singular) is minimised at $x = \funds$.
\end{lemma}
		
\begin{proof}
	With the change of variables $x \to c^{1/2} x$, and with $\rho \dfn c^{1/2} \funds$, we need to show that the function $\Real^{I \times K} \ni y \mapsto \norm{(y^* y)^{1/2} (\rho^* y)^{-1} \eta}^2_F$ is minimised at $y = \rho$ over all $y \in \Real^{I \times K}$ such that $\rho^* y$ is non-singular. Noting that
	\[
	\norm{(y^* y)^{1/2} (\rho^* y)^{-1} \eta}^2_F = \trace(\eta^* (y^* \rho)^{-1} y^* y (\rho^* y)^{-1} \eta ) = \norm{y (\rho^* y)^{-1} \eta}^2_F,
	\]
	we need to show that the function
	\[
	\Real^{I \times K} \ni y \mapsto \norm{y (\rho^* y)^{-1} \eta}^2_F
	\]
	is minimised at $y = \rho$ over all $y \in \Real^{I \times K}$ such that $\rho^* y$ is non-singular.
	
	Set $p = \rho (\rho^* \rho)^{-1} \rho^* \in \Real^{I \times I}$, and $\id$ the identity $\Real^{I \times I}$ matrix. We note that $p$ is a symmetric projection matrix: $p^* = p$, and $p^2 = p^* p = p$. Any $z \in \Real^{I \times K}$ is decomposed as $z = p z + (\id - p) z$. We observe that
	\[
	\inner{p z}{(\id - p) z}_F = \trace(z^* p^* (\id - p) z) = 0,
	\]
	since $p^* (\id - p) = 0$, which means that $p z$ and $(\id - p) z$ are $\inner{\cdot}{\cdot}_F$-orthogonal. Furthermore,
	\[
	p y (\rho^* y)^{-1} = \rho (\rho^* \rho)^{-1} \rho^* y (\rho^* y)^{-1} = \rho (\rho^* \rho)^{-1}.
	\]
	The previous observations give, for all $y \in \Real^{I \times K}$ such that $\rho^* y$ is invertible, 
	\begin{align*}
	\norm{y (\rho^* y)^{-1} \eta}^2_F &= \norm{p y (\rho^* y)^{-1} \eta}^2_F + \norm{(\id - p) y (\rho^* y)^{-1} \eta}^2_F \geq \norm{p y (\rho^* y)^{-1} \eta}^2_F \\
	&= \norm{\rho (\rho^* \rho)^{-1} \eta}^2_F,
	\end{align*}
	which is what we wanted to show.
\end{proof}

\subsection{An auxiliary lemma for \S\ref{SS:loss}}

This appendix provides a justification for the inequality in \eqref{eq:201121}.  We consider here a slightly more general framework than the one in \S\ref{SS:loss}. To motivate this setup, define the $(I \times I)$-matrix-valued-process $P := \funds (\funds^* \funds)^{-1} \funds^*$, which consists of orthogonal projections 
on the subspace generated by $f$. Consider an investment universe whose available returns are described by $\d R_P \dfn P^* \d R = P \d R$. While there are $I$ investment opportunities, where typically $|I| > |K|$, the investment universe is exactly the same as the one restricted to  the funds $\funds$.  In other words, being able to invest in a market whose returns are described by $\d R_P$ is economically equivalent to being able to invest only in the funds, whose returns are described by $\d R_\funds = \funds^* \d R$.

Although more abstract, this alternative point of view, which describes a restricted set of investment opportunities by an orthogonal projection matrix of rank $K$ (instead of a family of $K$ funds), is mathematically quite convenient.  Moreover, it also provides a more powerful tool to describe the investment restrictions. Instead of listing the potential funds, it now suffices to specify the ($\cF$--adapted) matrix (process) $P$. This framework is more flexible; for example, it incorporates the setup of a fluctuating number of funds (corresponding to the rank of $P$) in which one is allowed to invest.

As already mentioned, in this alternative way of modelling a restricted investment universe, the available returns are described by $\d R_P \dfn P \d R$ with covariance differential $\d C_{PP} \dfn P (\d C) P$.  This yields the $\cG$-portfolio $\numg_P \dfn (\d C_{PP})^{\dagger} P (\d C) \numg$ 
 that maximises growth among all investment opportunities restricted to $P$, where $(\d C_{PP})^{\dagger}$ formally denotes the inverse of $\d C_{PP}$. (To make this precise, we interpret $\d C_{PP}$  as a linear isomorphism on the subspace associated with the projection $P$; hence having an inverse on that subspace.) Similarly, the $\cF$-growth-optimal portfolio equals $\numf_P \dfn (\d C_{PP})^{\dagger} P (\d C) \numf$. 
 With $\d G_P$ and $\d F_P$ now denoting the differential of the maximal archivable growth under the information flows $\cG$ and $\cF$ in the restricted investment universe, from \eqref{eq:200126.2} one obtains 
\begin{align*}
	\EF [\d G_P] - \d F_P &=   \frac{1}{2} \trace \left(  (\d C_{PP})^{\dagger} P (\d C)  \varnu (\d C) P    (\d C_{PP})^{\dagger}   (\d C_{PP})\right) 
	\\
	&= 	\frac{1}{2} \trace  \left((\d C_{PP})^{\dagger} P (\d C) \varnu (\d C) P  \right).
\end{align*}

 We now want to argue 
  \begin{align} \label{eq:201121.2}
	\frac{1}{2} \trace  \left((\d C_{PP})^{\dagger} P (\d C)  \varnu (\d C) P  \right) \leq \frac{1}{2} \trace  (\varnu \d C   ).	
\end{align}
 Indeed, this corresponds exactly to the inequality in \eqref{eq:201121} when $P := \funds (\funds^* \funds)^{-1} \funds^*$ since in this case $\d G_P = \d G_\funds$ and $\d F_P = \d F_\funds$. The inequality \eqref{eq:201121.2} implies that a restriction of the investment universe yields a smaller loss in the respective optimal growth when one goes from the larger to the smaller information flow.  The next lemma now yields \eqref{eq:201121.2}, hence also the inequality in \eqref{eq:201121}.

\begin{lemma} \label{lem:error_reduction}
Consider a symmetric nonnegative definite matrix $c \in \Real^{I \times I}$ and an orthogonal projection matrix $p \in \Real^{I \times I}$ (that is, $p^2 = p = p^*$). With $c_{pp} \dfn p^* c p = p c p$, let $c_{pp}^{\dagger}$ denote the inverse of $c_{pp}$ when viewed as a linear mapping on the subspace associated with $p$. Then, it holds that  $c c_{pp}^{\dagger}  c \leq c$ in the order of nonnegative definite matrices.
\end{lemma}

\begin{proof}
	We may assume that $c$ is invertible; otherwise, just work on the subspace generated by the range of $c$.
	To ease notation in the course of the proof, set $c_p \dfn c_{pp}$. Then, we need to show that $c_p^{\dagger}  \leq c^{-1}$. Define the orthogonal projection matrix $q \dfn \id - p$, and note that $q p = 0 = p q$. Furthermore, set $h \dfn c^{-1}$,  $h_p \dfn p hp$, and $h_q \dfn qhq = (\id-p) h(\id-p)$, and note that $c h = \id$ gives  in particular  $c_p h_p + p c q h p = p$ and  $c_p h q + p c q h_q= 0$. The second equation gives $p c q = - c_p h q h_q^{\dagger} = - c_p h h_q^{\dagger}$; plugging this back in to the first equation, we obtain  $c_p h_p = p + c_p h h_q^{\dagger} h p$. This last equation is equivalent to  $h_p = c_p^{\dagger}  + p h h_q^{\dagger} h p$, giving  $c_p^{\dagger} = h_p - p h h_q^{\dagger} h p$.

	For $x \in \Real^I$, we have
	\[
	x^* (h - c_p^{\dagger}) x = 2 x^* p h q x + x^* q h q x + x^* p h h_q^{\dagger} h p x.	
	\]
	With  $z \dfn q x$ and $w \dfn h p x$, we need to show
	\[
		2 w^* z + z^* h z + w^* h_q^{\dagger} w \geq 0.
	\]
	Define now  $y \dfn q w$. Since  $p z = 0$, we have $z^* h z = z^* h_q z$ and  $w^* z = y^* z$. Since $h_q^{\dagger} p = 0$,  we have  $w^* h_q^{\dagger} w = y^* h_q^{\dagger} y$. Therefore, we need to show that $2 y^* z + z^* h_q z + y^* h_q^{\dagger} y \geq 0$. Since everything now lies on the subspace associated with the projection $q = 1-p$, we have
	\[
		2 y^* z + z^* h_q z + y^* h_q^{\dagger} y = \norm{h_q^{1/2} z + (h_q^{\dagger})^{1/2} y}^2,	
	\]
	which is nonnegative, establishing the claim.
\end{proof}

\subsection{Auxiliary results for Section~\ref{S:Shrinkage}}
The following lemma and remarks are used to describe the optimal shrinkage in Section~\ref{S:Shrinkage}.
\begin{lemma} \label{lem:shrink_multi_d}
	For given $z \in \Real^I$ and symmetric nonnegative definite matrix $h \in \Real^{I \times I}$, the problem
	\[
	\arg \min_{y \in \Real^I} \left( \frac{1}{4} \norm{y - z}^4 + \norm{h^{1/2} y}^2 \right)\]
	has a unique solution, which is $y = z$ if $z \in \ker(h)$ and  $y = \left( \id + b^{-1} h \right)^{-1} z$ if $z \notin \ker(h)$. Here $b \in (0, \norm{z}^2/2)$ is the unique solution\footnote{When $z \notin \ker(h)$, note that $(0, \infty) \ni b \mapsto \norm{ ( h + b \id )^{-1} h z}^2 \in (0, \infty)$ is continuous and strictly decreasing, with limit equal to zero as $b \uparrow \infty$. Furthermore, at $b = \norm{z}^2/2$, the value of the above function is strictly less than $\norm{z}^2/2$.} of the one-dimensional equation
	\begin{align}\label{eq:201121.3}
	\frac{1}{2} \norm{ \left( h + b \id \right)^{-1} h z}^2 = b.	
	\end{align}
\end{lemma}

\begin{proof}
	This strictly convex problem in $y$ has a unique minimiser. First order conditions give
	\[
	\norm{y - z}^2 (y - z) + 2 h y = 0.	
	\]
	If $z \in \ker (h)$, then $y = z$. Otherwise, with $b \dfn (1/2) \norm{y - z}^2 \in (0, \infty)$, we have $y = \left( \id + b^{-1} h \right)^{-1} z$. In order to identify $b > 0$, we have
	\[
	2 b = \norm{y - z}^2 = \norm{ \left( \id - \left( \id + b^{-1} h \right)^{-1} \right) z}^2 = \norm{ \left( h + b \id \right)^{-1} h z}^2 < \norm{z}^2,	
	\]
	where we have used the identity $\id - \left( \id + b^{-1} h \right)^{-1} = \left( h + b \id \right)^{-1} h$.
\end{proof}

\begin{remark} \label{rem:on_shrink_multi_d}
	In the context of Lemma~\ref{lem:shrink_multi_d}, let $y$ denote the optimal solution and assume that $y \neq z$. Then  the first-order condition in the proof of  Lemma~\ref{lem:shrink_multi_d} yields
	\[
		y^* (z - y) = y^* b^{-1} h y = b^{-1} \norm{h^{1/2} y}^2  = 2  
		\frac{\norm{h^{1/2} y}^2 }{ \norm{y - z}^2}.
	\]
	 Therefore,
	\[
	\norm{z}^2 = \norm{y}^2 + \norm{y - z}^2 + 4 \frac{\norm{h^{1/2} y}^2}{\norm{y - z}^2} = \norm{y}^2 + 4 \frac{(1/4) \norm{y - z}^4 + \norm{h^{1/2} y}^2}{\norm{y - z}^2}	.
	\]
	In particular, $\norm{y}^2 \leq \norm{z}^2$.
\end{remark}

The following remark yields a numerical algorithm to determine a solution $b$ to the equation in \eqref{eq:201121.3}.

\begin{remark} \label{rem:on_fixed_point}
	Recall the notation of Lemma~\ref{lem:shrink_multi_d}. Upon writing $h = \sum_{i \in I} s_i v_i v_i^*$	for the eigenvalues $(s_i; i \in I) \in \Real_+^I$ and corresponding eigenvectors $(v_i; i \in I) \in \Real^{I \times I}$ of $h$, we have
	\[
		f(b) = \frac{1}{2} \norm{ \left( h + b \id \right)^{-1} h z}^2 = \frac{1}{2} \sum_{i \in I} \left( \frac{s_i}{s_i + b}  \right) ^2 (v_i^* z)^2.
	\]
	Then $f(b) = b$ is a simple non-linear equation to be solved numerically, as discussed below. 
	
	Consider a strictly decreasing convex function $f: [0, \infty) \mapsto (0, \infty)$ with $x_0 \dfn f(0) > 0$. Then, the fixed point equation $f(x) = x$ will have a unique solution, which in fact has to satisfy $0 < x < x_0$. One can use a simple fixed-point algorithm. Set $\bbL \dfn \{ x \in [0, x_0] : x \leq f(x) \}$ and $\bbU \dfn \{ x \in [0, x_0]  : f(x) \leq x \}$, noting that the interval $\bbL$ lies on the left of the interval $\bbU$. Note that $x_0 \in \bbU$. We shall define inductively a decreasing $\bbU$-valued sequence $(x_n)_{n \in \Natural}$ that will converge to the fixed point. Given $x_{n-1} \in \bbU$ and $w_{n-1} \dfn f(x_{n-1})\in \bbL$ (the last is a definition of $w_{n-1}$, and the fact that $w_{n-1} \in \bbL$ holds by the fact that $f$ is decreasing), convexity of $f$ implies that
	\[
	f(x) \leq f(w_{n-1}) + (x - w_{n-1}) \frac{f(x_{n-1}) - f(w_{n-1})}{x_{n-1} - w_{n-1}}, \quad w_{n-1} \leq x \leq x_{n-1}.
	\]
	Therefore, if $x_n \in [w_{n-1}, x_{n-1}]$ satisfies the equality
	\[
		f(w_{n-1}) + (x_n - w_{n-1}) \frac{f(x_{n-1}) - f(w_{n-1})}{x_{n-1} - w_{n-1}} = x_n,	
	\]
	we have $f(x_n) \leq x_n$, i.e., $x_n \in \bbU$. Explicitly, we have
	\[
	x_n = \frac{(x_{n-1} - w_{n-1}) f(w_{n-1}) + (f(w_{n-1}) - w_{n-1}) w_{n-1}}{(x_{n-1} - w_{n-1}) + (f(w_{n-1}) - w_{n-1})}.	
	\]
	Setting $w_n \dfn f(x_n) \in \bbL$, we have $w_{n-1} \leq w_n$; in fact, with
	\[
	\eta_{n-1} \dfn \frac{f(w_{n-1}) - w_{n-1}}{(x_{n-1} - w_{n-1}) + (f(w_{n-1}) - w_{n-1})} \in [0, 1],	
	\]
	the above definition of $x_n$ gives
	\[
	x_n - f(x_n) = x_n - w_n \leq x_n - w_{n-1} = \eta_{n-1} (x_{n-1} - w_{n-1}) = \eta_{n-1} (x_{n-1} - f(x_{n-1}))	
	\]
	We have shown that $(x_n)_{n \in \Natural}$ is decreasing, and note that $x_n - f(x_n)$ also decreases in $n$. The limit $x_\infty$ exists, as does $w_\infty$, and we have $f(x_\infty) = w_{\infty} \leq x_\infty$. The above iteration also implies that, if $x_\infty - f(x_\infty) > 0$, then $f(w_\infty) = w_\infty$. But then, since $x_\infty \leq f(w_\infty) = w_\infty$, we would have $x_\infty = f(x_\infty)$. We conclude that $w_\infty = x_\infty$, and that this is the (unique) fixed point.

	For further improvement, we can also choose $w_n$ in a better way, provided we also calculate $f'(x_n)$, which in our case is not a problem. Indeed, convexity again gives
	\[
	f(w) \geq f(x_n) + f'(x_n) (w - x_n), \quad 0 \leq w \leq x_0;	
	\]
	therefore, if $w_n$ satisfies $f(x_n) + f'(x_n) (w_n - x_n) = w_n$, we have $w_n \in \bbL$. Solving this, we obtain
	\[
	w_n = \frac{1}{1 - f'(x_n)} f(x_n) + \frac{- f'(x_n)}{1 - f'(x_n)} x_n,	
	\]
	a convex combination between $f(x_n)$ and $x_n \geq f(x_n)$, which gives $w_n \geq f(x_n)$. Note that this step is just the Newton-Raphson method for solving $f(x) - x = 0$; what convexity gives us is a handle on the improvement on the previous iteration. 
\end{remark} 

The following remark is a sanity check for the computations in \S\ref{SS:UniformShrink}.

\begin{remark} \label{rem:reduction_multi_d}
	In the context of Lemma \ref{lem:shrink_multi_d}, suppose that $h$ is a constant multiple $s > 0$ of the identity matrix; i.e., $h = s \, \id$. In this case, $b$ (rather, $b/s$) satisfies the 3rd order equation
	\[
		\left(\frac{b}{s}\right)^3 + 2 \left(\frac{b}{s}\right)^2 + \frac{b}{s} - \frac{\norm{z}^2}{2s} = 0.	
	\]
	Substituting $a \dfn b/(b+s)$ then gives $y = a z$ and $b/s = a / (1-a)$.
	A short computation yields that $a$ satisfies $- (\norm{z}^2/s) (1-a)^3 + 2 a = 0$, which leads indeed to \eqref{eq:200628.3}, with
	\[
	\psi = \left( \frac{3}{2} \right)^2 \frac{\norm{z}^2}{s} .	
	\]
\end{remark}

\section{Online appendix: Extensions of the empirical study} \label{A:data}
We now extend the empirical analysis of Section~\ref{S:empirics} to four other international equity markets: UK, Germany, Italy, and Australia.  The selection of these countries was purely based on the availability of daily observations of a total return index (i.e., incorporating dividends) and risk-free rates over a sufficiently long time period (at least 40 years). 

\begin{figure}
	\includegraphics[width=1\textwidth]{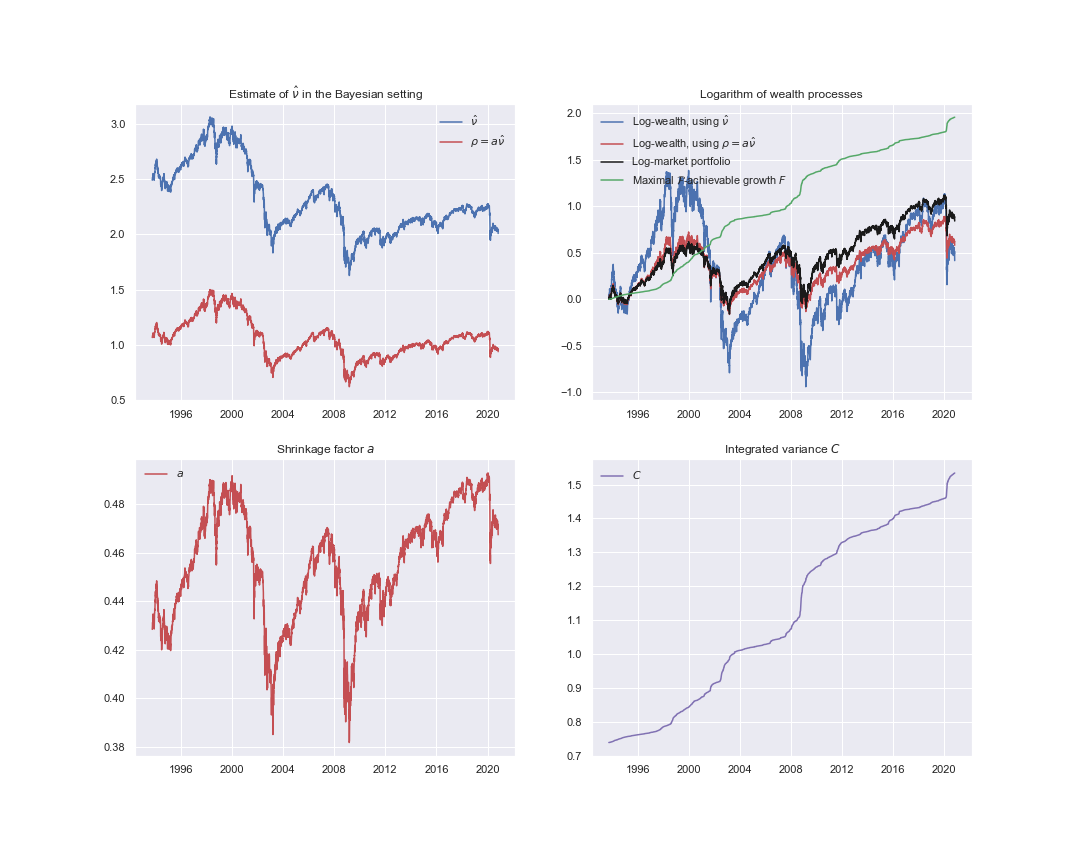}
	\caption{UK data.}
	\label{F:UK}
\end{figure}

\begin{figure}
	\includegraphics[width=1\textwidth]{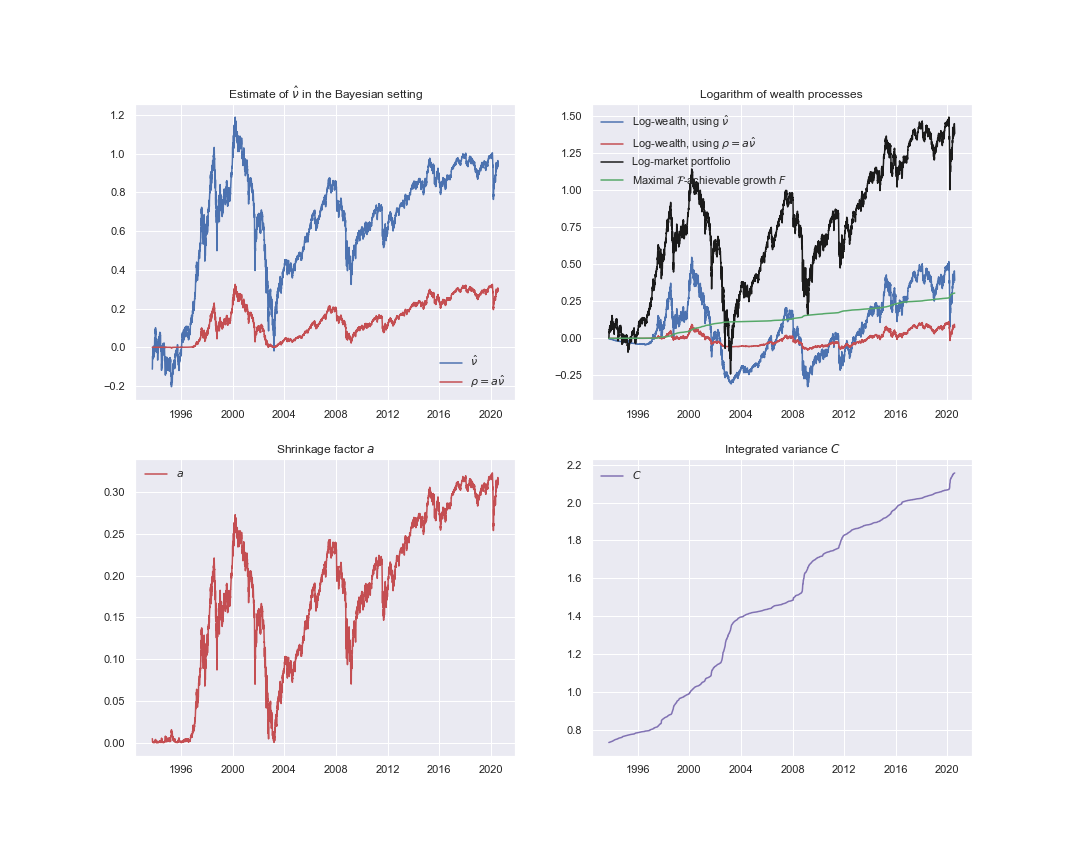}
	\caption{German data.}
	\label{F:Germany}
\end{figure}

\begin{figure}
	\includegraphics[width=1\textwidth]{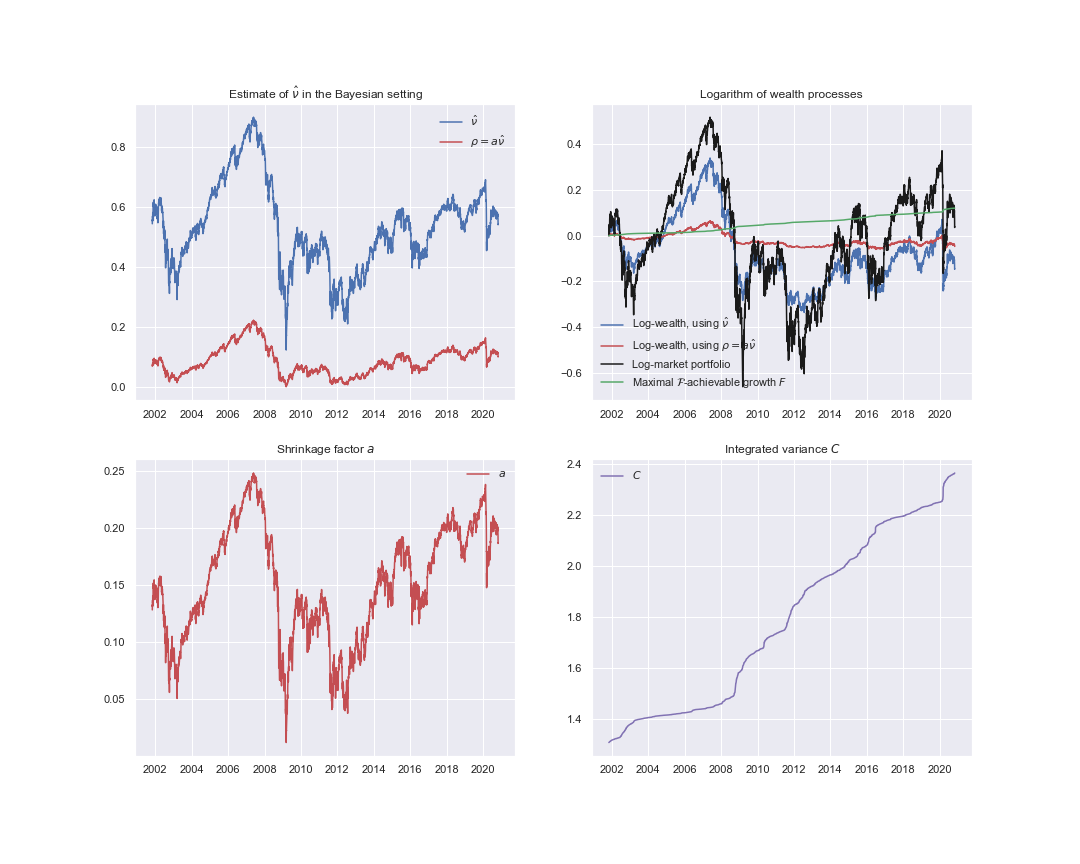}
	\caption{Italian data.}
	\label{F:Italy}
\end{figure}

\begin{figure}
	\includegraphics[width=1\textwidth]{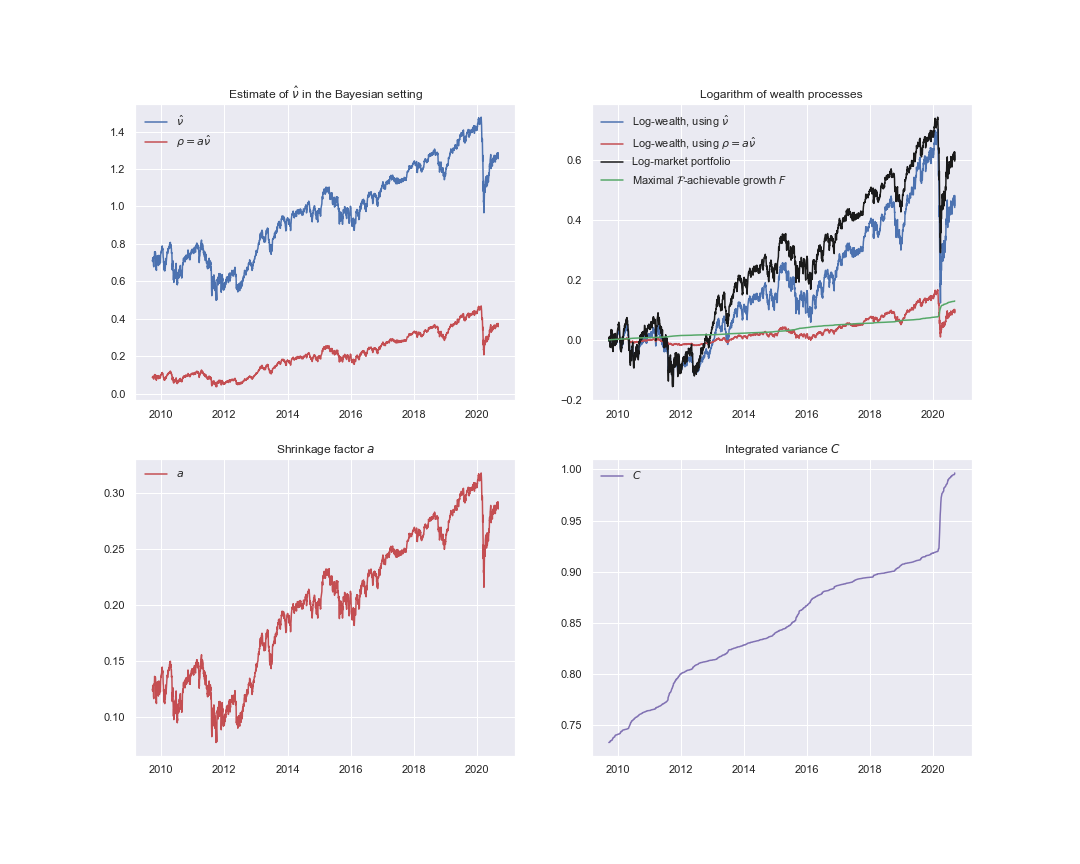}
	\caption{Australian data.}
	\label{F:Australia}
\end{figure}

The figures below summarise the empirical findings. We order them according to the length of the available time series of factor returns:  UK (Figure~\ref{F:UK}), Germany  (Figure~\ref{F:Germany}), Italy  (Figure~\ref{F:Italy}), and Australia (Figure~\ref{F:Australia}).    The four panels are exactly as for the US market; see Section~\ref{S:empirics}. Recall that we start plotting the time series  7500 trading days (i.e., about 30 years) after the first data are available.   Since the availability of  various time series (returns and risk-free rates) various across the different global markets the different figures have different starting dates.   

The UK market returns correspond to the UK FTSE All-Share Return index of Global Financial Data.  Daily data are available from January 1965 and go to October 2020.   The risk-free rates used are the 3-month Treasury Bill Yields, again obtained from Global Financial Data.

The German market returns correspond to the German  total return index of Refinitiv Datastream.  Data are available from January 1965 to July 2020.  The risk-free rates are the Bundesbank Lombard rates until December 1998, and the ECB's deposit facility rate afterwards. These rates are downloaded from Deutsche Bundesbank. We update the rates at the beginning of each month.  We also obtained the Germany CDAX Total Return Index of Global Financial Data.
For this index, daily data are available from January 1970. Comparing the two datasets on the restricted time span did not show any significant differences, hence we decided to use the Refinitiv Datastream data, for which we have five more years of data.

The Italian market returns correspond to the UK MIB Return index of Global Financial Data.  Daily data are available from January 1973 and go to October 2020.   The risk-free rates used are the 3-month Treasury Bill Yields, again obtained from Global Financial Data.

The Australian market returns correspond to the Australian  total return index of Bloomberg.  As risk-free interest rate we use the Cash Rate (AONIA) of the Reserve Bank of Australia. We obtained its time series from Refinitiv Datastream (which provides a longer time series than the one provided by the Reserve Bank of Australia).  Although the Bloomberg return index is very long we could only get risk-free rates from February 1980 onwards. Hence, we start the analysis then.

\bibliographystyle{apalike}
\bibliography{library}

\end{document}